\documentclass[11pt,a4paper]{article}
\usepackage{color}
\usepackage[latin1]{inputenc}
\usepackage[T1]{fontenc}
\usepackage[english]{babel}
\usepackage{amsfonts}
\usepackage{amssymb}
\usepackage{graphicx}
\usepackage{dsfont}
\usepackage{epsfig}
\usepackage{ae}
\usepackage{hyperref}
\usepackage{harvard}
\usepackage{epstopdf}
\usepackage{amsmath}
\usepackage{setspace}
\usepackage{amsthm}
\usepackage{appendix}
\usepackage{float}

\renewcommand{\proof}{{\noindent \it Proof. }}
\newtheorem{theo}{Theorem}
\newtheorem{pr}{Proposition}
\newtheorem{lem}{Lemma}
\newtheorem{co}{Corollary}
\newtheorem{defi}{Definition}
\theoremstyle{remark}
\newtheorem{re}{Remark}

\newcommand{\be}{\begin{eqnarray}}
\newcommand{\ee}{\end{eqnarray}}
\newcommand{\by}{\begin{eqnarray*}}
\newcommand{\ey}{\end{eqnarray*}}
\newcommand{\bt}{\begin{theo}}
\newcommand{\et}{\end{theo}}
\newcommand{\bl}{\begin{lem}}
\newcommand{\el}{\end{lem}}
\newcommand{\bc}{\begin{co}}
\newcommand{\ec}{\end{co}}
\newcommand{\eex}{\end{exa}\vspace{-3mm}}
\newcommand{\br}{\begin{re}}
\newcommand{\er}{\end{re}\vspace{-3mm}}
\renewcommand{\geq}{\geqslant}
\renewcommand{\leq}{\leqslant}
\renewcommand{\ge}{\geqslant}
\renewcommand{\le}{\leqslant}

\citationstyle{agsm}
\allowdisplaybreaks
\renewcommand{\tilde}{\widetilde}
\input{tcilatex}
\begin{document}

\title{\textbf{\Large Rationalizing Investors Choice}}
\author{Carole Bernard$^{\dag }$\thanks{%
Corresponding author: \texttt{c3bernar@uwaterloo.ca}. C. Bernard
acknowledges support from the Natural Sciences and Engineering Research
Council of Canada. S. Vanduffel gratefully acknowledges the financial
support of the BNP Paribas Fortis Chair in Banking. We thank participants in
the seminar in Geneva, the workshops on Quantitative Behavioral Finance in
St Gallen (September 2012) and Waterloo (April 2013), and in the annual
meetings of AFFI in Lyon (May 2013), ARIA in Washington DC (August 2013) and
of EGRIE in Paris (September 2013).}, Jit Seng Chen$^{\ddag \diamond}$ and
Steven Vanduffel$^{\diamond }$ \\
\noindent \textit{\small $^\dag$Department of Statistics and Actuarial
Science, University of Waterloo, Canada.}\\
\noindent \textit{\small $^\ddag$Gilliland Gold Young Consulting, Toronto,
Canada.}\\
\noindent \textit{\small $^\diamond$Faculty of Economic, Political and
Social Sciences}\\
\textit{\small and Solvay Business School, Vrije Universiteit Brussel,
Belgium.} }
\date{\today }
\maketitle

\begin{abstract}
Assuming that agents' preferences satisfy first-order stochastic dominance,
we show how the Expected Utility paradigm can rationalize all optimal
investment choices: the optimal investment strategy in any behavioral
law-invariant (state-independent) setting corresponds to the optimum for an
expected utility maximizer with an explicitly derived concave non-decreasing
utility function. This result enables us to infer the utility and risk
aversion of agents from their investment choice in a non-parametric way. We
relate the property of decreasing absolute risk aversion (DARA) to
distributional properties of the terminal wealth and of the financial
market. Specifically, we show that DARA is equivalent to a demand for a
terminal wealth that has more spread than the opposite of the log pricing
kernel at the investment horizon.
\end{abstract}

\onehalfspacing




\newpage 

\section{Introduction}

First suggested by Bernoulli \citeyear{B} and rigorously introduced in the
economic literature by von Neumann and Morgenstern \citeyear{VNM}, Expected
Utility Theory (EUT) has for decades been the dominant theory for making
decisions under risk. Nonetheless, this framework has been criticized for
not always being consistent with agents' observed behavior (e.g., the
paradox of Allais \citeyear{allais53}, Starmer \citeyear{starmer}). In
response to this criticism, numerous alternatives have been proposed, most
notably dual theory (Yaari \citeyear{yaari}), rank-dependent utility theory
(Quiggin \citeyear{quiggin93}) and cumulative prospect theory (Tversky and
Kahneman \citeyear{cptKT92}). These competing theories differ significantly,
but all three typically satisfy first-order stochastic dominance (FSD).
Indeed, many economists consider violation of this property as grounds for
refuting a particular model; see for example Birnbaum \citeyear{Birnbauma},
Birnbaum and Navarrette \citeyear{Birnbaumb}, and Levy \citeyear{Levy} for
empirical evidence of FSD violations. Recall also that although the original
prospect theory of Kahneman and Tversky \citeyear{ptKT79} provides
explanations for previously unexplained phenomena, it violates FSD. To
overcome this potential drawback, Tversky and Kahneman \citeyear{cptKT92}
developed cumulative prospect theory.

In the presence of a continuum of states, we show that the optimal portfolio
in any behavioral theory that respects FSD can be rationalized by the
expected utility setting, i.e., it is the optimal portfolio for an expected
utility maximizer with an explicitly known concave utility function. This
implied utility function is unique up to a linear transformation among
concave functions and can thus be used for further analyses of preferences,
such as to infer the risk aversion of investors. 

A surprising feature is that we only assume that the preferences respect
FSD, which contrasts to earlier results on the rationalization of investment
choice under expected utility theory. Dybvig (\citeyear{dybvigJoB}, Appendix
A), Peleg and Yaari \citeyear{PY75} and Zilcha and Chew %
\citeyear{zilcha1990invariance}, among others, have worked on this problem
assuming that preferences preserve second-order stochastic dominance (SSD).
However, being SSD-preserving is quite a strong assumption, and while
consistency with FSD is inherent, and even enforced, in most decision
theories, this is not readily the case for SSD. For instance, rank dependent
utility theory satisfies FSD but not SSD (Chew, Karni and Safra %
\citeyear{hong1987risk}, Ryan \citeyear{ryan2006risk}), and the same holds
true for cumulative prospect theory (see e.g., Baucells and Heukamp %
\citeyear{baucells2006stochastic}). Thus, our results show that for any
agent behaving according to the cumulative prospect theory (i.e., for a
``CPT investor'') there is a corresponding expected utility maximizer with
concave utility purchasing the same optimal portfolio, even if the CPT
investor can exhibit risk seeking behavior with respect to losses. This
approach, however, is not intended to dispense with alternative models to
expected utility theory, as they have been developed mainly to compare
gambles and not to deal with optimal portfolio selection per se.

Our results are rooted in the basic insight that under some assumptions, the
marginal utility at a given consumption level is proportional to the ratio
of risk-neutral probabilities and physical probabilities (Nau and Mccardle %
\citeyear{nau1991arbitrage}, Duffie \citeyear{duffie2010dynamic}, Pennacchi %
\citeyear{pennacchi2008theory}). At first, it then seems obvious to infer a
(concave) utility function and the risk aversion from the optimal
consumption of the investor. However, the characterization that the marginal
utility is proportional to the pricing kernel at a given consumption level
is valid only if the utility is differentiable at this consumption level.
This observation renders the rationalization of investment choices by the
expected utility theory non-trivial, as there are many portfolios for which
the implied utility is not differentiable at all consumption levels, such as
the purchase of options or capital guarantee products. Furthermore, in a
discrete setting (with a finite number of equiprobable states) there are
many utility functions that are consistent with optimal consumption. In this
context, Peleg and Yaari \citeyear{PY75} give one potential implied utility,
but there are many others. In the presence of a continuum of states, when
the pricing kernel is continuously distributed, we are able to derive the
unique (up to a linear transformation) concave utility function that is
implied by the optimal consumption of any investor who respects FSD.

The proof of our main results builds on Dybvig's %
\citeyear{dybvigJoB,dybvigRFS} seminal work on portfolio selection. Instead
of optimizing a value function, Dybvig \citeyear{dybvigJoB} specifies a
target distribution and solves for the strategy that generates the
distribution at the lowest possible cost. \footnote{%
It may indeed be more natural for an investor to describe her target
distribution of terminal wealth instead of her utility function. For
example, Goldstein, Johnson and Sharpe {\citeyear {GJS2}} discuss how to
estimate the distribution at retirement using a questionnaire. 
The pioneering work in portfolio selection by Markowitz \citeyear{Marko} and
Roy \citeyear{Roy} is based solely on the mean and variance of returns and
does not invoke utility functions. Black \citeyear{Black} calls a utility
function \textquotedblleft a foreign concept for most
individuals\textquotedblright\ and states that \textquotedblleft instead of
specifying his preferences among various gambles the individual can specify
his consumption function". 
} Our results are also related to earlier work on revealed preferences by
Afriat \citeyear{A} and Varian \citeyear{varian1982,varian2006}, whose
analyses aim at understanding the preferences of a consumer who has to
choose among bundles of goods, given a budget constraint. Their original
problem can be formulated as follows. Given a set of observations of prices
and quantities for a finite number of goods observed over some period of
time, is it possible to find a utility function that rationalizes the
observed portfolio choices? That is, at any time $t$, the utility obtained
by the portfolio of goods (observed at time $t$) is at least as good as any
other available portfolio at time $t$ for this utility. If the set of
observed price-quantity pairs satisfies some inequalities (called cyclical
consistency), Afriat \citeyear{A} proves the existence of an increasing,
continuous and concave utility function. 
Here, we analyze the idea of inferring preferences for consumers who are
investing in the financial market. We show that if their portfolio satisfies
some conditions, then it can be rationalized by expected utility theory with
a non-decreasing and concave function. Although our analysis appears to be a
natural extension (see Theorem \ref{thD} and Remark \ref{rkk}), it differs
fundamentally from Afriat's work in a few aspects. First, Afriat \citeyear{A}%
, Varian \citeyear{varian1982,varian2006} and most of the subsequent work on
revealed preferences has been done in a \textquotedblleft
certainty\textquotedblright\ framework (utility setting), whereas, in the
financial market, the future consumption for each unit invested is unknown
(expected utility setting). 
Secondly, Afriat's \citeyear{A} approach relies on an ex-post analysis of a
sequence of observed consumptions in time, whereas our analysis studies the
ex-ante properties of a future consumption bundle in a one-period setting.
Finally, we assume that there is an infinite number of states in which it is
possible to invest, whereas the analyses of Afriat \citeyear{A} and Varian %
\citeyear{varian1982,varian2006} are formulated with a finite number of
states. The assumption that we make is natural in the context of optimal
portfolio selection problems. It allows to obtain the uniqueness of the
implied concave utility and to be able to use this inferred utility to
estimate risk aversion, for instance.

Inference of risk preferences from observed investment behavior has also
been studied by Sharpe \citeyear{Sha}, Sharpe, Goldstein and Blythe %
\citeyear{SGB}, Dybvig and Rogers \citeyear{DR} and Musiela and
Zariphopoulos \citeyear{MZ}.\footnote{%
Under some conditions, Dybvig and Rogers \citeyear{DR} and Musiela and
Zariphopoulos \citeyear{MZ} infer utility from dynamic investment decisions.
Our setting is static and well adapted to the investment practice by which
consumers purchase a financial contract and do not trade afterwards.} Sharpe %
\citeyear{Sha} and Sharpe et al. \citeyear{SGB} assume a static setting and
rely on Dybvig's \citeyear{dybvigRFS} results to estimate the coefficient of
constant relative risk aversion for a CRRA utility based on target
distributions of final wealth. 

In this paper, we establish a link between expected utility theory (EUT) and
all other theories that respects FSD. 
This connection can be used to estimate the agents' utility functions and
risk aversion coefficients in a non-expected utility setting. Our approach
in doing so is non-parametric and is based solely on knowledge of the
distribution of optimal wealth and of the financial market. This is in
contrast to traditional approaches to inferring utility and risk aversion,
which specify an exogenous parametric utility function in isolation of the
market in which the agent invests and then calibrate this utility function
using laboratory experiments and econometric analysis of panel data. 

It is widely accepted that the Arrow-Pratt measure of absolute risk aversion
is decreasing with wealth. This feature - i.e., decreasing absolute risk
aversion (DARA) - is often the motivation for using the CRRA utility instead
of the exponential utility to model investors' preferences. 
In this paper, we show that the DARA property is completely characterized by
a demand for final wealth $W$ that exhibits more spread than a certain
market variable (which we define in a precise sense in the paper). 
Our characterization of DARA can be used to empirically test DARA
preferences based on observed investment decisions.

\section{Introductory Example\label{S0}}

Throughout this paper, we consider agents with law-invariant and
non-decreasing preferences,\footnote{%
This assumption is present in most traditional decision theories including
expected utility theory (von Neumann and Morgenstern \citeyear{VNM}),
Yaari's dual theory (Yaari \citeyear{yaari}), the cumulative prospect theory
(Tversky and Kahneman \citeyear{cptKT92}) and rank dependent utility theory
(Quiggin \citeyear{quiggin93}).} $V(\cdot )$. We say that $V(\cdot )$ is 
\textit{non-decreasing} if, for consumptions $X$ and $Y$ satisfying $X\leq Y$%
, one has that $V(X)\leq V(Y).$ We say that $V(\cdot )$ is \textit{%
law-invariant} if $X\sim Y$ implies that $V(X)=V(Y),$ where
\textquotedblleft $\sim $\textquotedblright\ reflects equality in
distribution. This is often referred as a \textquotedblleft
state-independent\textquotedblright\ set of preferences. We also assume that
the agent's initial budget is finite.

In this section, we present an example in order to introduce the notation
and to explain in a simplified setting (a space with a finite number of
equal probable states) why a distribution of terminal wealth can always be
obtained as the optimum of the maximization of expected utility for a risk
averse agent. We will also show the limitations of this discrete setting and
how it fails to identify \textit{the} implied concave utility function and
implied risk aversion of the investor.

The introductory example takes place in a finite state space $\Omega
=\left\{ \omega _{1},\omega _{2},...,\omega _{N}\right\} $ consisting of $N$
equiprobable states (with probability $\frac1N$) at some terminal time $T$.
Denote by $\frac{\xi (\omega _{i})}{N}$ the initial (positive) cost at time
0 of the Arrow-Debreu security that pays one unit in the $i^\text{th}$
state, $\omega _{i}$, at time $T$ and zero otherwise. Let us call $\xi
:=\left( \xi_1,\xi_2,...,\xi_N\right) $ the pricing kernel where $%
\xi_i:=\xi(\omega_i)$. It is clear that any state-contingent consumption $%
X:=(x_1,x_2,...,x_N)$ (with $x_i:=X(\omega_i)$) at time $T$ writes as a
linear combination of the $N$ Arrow--Debreu securities.

The optimal investment problem of the agent with preferences $V(\cdot )$ is
to find the optimal consumption $X^{\ast }$ by solving the optimization
problem, 
\begin{equation}
\max_{X\ |\ E[\xi X]=X_{0}}V(X),  \label{RE}
\end{equation}%
where the budget constraint $E[\xi X]=\frac{1}{N}\sum_{i=1}^{N}\xi
_{i}x_{i}=X_{0}$ reflects that the agent's initial wealth level is $X_{0}$.
We assume that an optimum $X^{\ast }$ to Problem \eqref{RE} exists (it is
always the case when restricting to non-negative consumptions). Denote by $%
x_{i}^{\ast }:=X^{\ast }(\omega _{i})$. Observe then that $X^{\ast }$ and $%
\xi $ must be anti-monotonic\footnote{%
This observation appeared in Peleg and Yaari \citeyear{PY75} (as the
principle of decreasing willingness in Theorem 1) and in Dybvig %
\citeyear{dybvigJoB,dybvigRFS}.}; in other words, the outcomes for $X^{\ast
} $ are ordered in reverse of the ones for $\xi $, or 
\begin{equation}
\forall \omega ,\omega ^{\prime }\in \Omega ,\quad \left( \xi (\omega )-\xi
(\omega ^{\prime })\right) \left( X^{\ast }(\omega )-X^{\ast }(\omega
^{\prime })\right) \leq 0.  \label{EQOM}
\end{equation}%
Let us prove \eqref{EQOM} by contradiction. To this end, assume that there
exist two states, $\omega _{i}$ and $\omega _{j}$, such that $\xi _{i}>\xi
_{j}$ and $x_{i}^{\ast }>x_{j}^{\ast }.$ Let $Y$ be another consumption such
that $Y(\omega )=X^{\ast }(\omega )$ for all $\omega \in \Omega \backslash
\left\{ \omega _{i},\omega _{j}\right\} $, and $Y(\omega _{i})=x_{j}^{\ast
}, $ $Y(\omega _{j})=x_{i}^{\ast }$. Since all states are equiprobable, $%
X\sim Y $ implies that $V(X)=V(Y)$ by the law-invariance of $V(\cdot )$.
However, $Y$ has a strictly lower cost, i.e., 
\begin{equation*}
E[\xi X^{\ast }]-E[\xi Y]=\frac{1}{N}(\xi _{i}-\xi _{j})(x_{i}^{\ast
}-x_{j}^{\ast })>0.
\end{equation*}%
Hence, $X^{\ast }$ cannot be optimal as $V(X^{\ast })<V(Z)$ in which $Z=Y+%
\frac{E[\xi X^{\ast }]-E[\xi Y]}{E[\xi ]}$ (note that $E[\xi Z]=X_{0}).$
Without loss of generality, using \eqref{EQOM}, we can thus assume that $%
x_{1}^{\ast }\leq x_{2}^{\ast }\leq ...\leq x_{N}^{\ast }$ and $\xi _{1}\geq
\xi _{2}\geq ...\geq \xi _{N}.$ For ease of exposition, we suppose in
addition that the inequalities are strict (see Peleg and Yaari %
\citeyear{PY75} for the most general case): 
\begin{equation*}
x_{1}^{\ast }<x_{2}^{\ast }<...<x_{N}^{\ast },\quad \quad \xi _{1}>\xi
_{2}>...>\xi _{N}.
\end{equation*}

\begin{pr}[Rationalizing Investment in a Discrete Setting]
\label{PR0} The optimal solution of \eqref{RE}, denoted by $X^{\ast }$ also
solves the maximum expected utility problem 
\begin{equation}
\max_{X\ |\ E[\xi X]=X_{0}}E[U(X)]  \label{RE2}
\end{equation}%
for any concave utility $U(\cdot )$ such that the left derivative\footnote{%
The results can also be written in terms of right derivative.} denoted by $%
U^{\prime }$ exists in $x_{i}^{\ast }$ for all $i$ and satisfies 
\begin{equation}
\forall i\in \{1,2,...,N\},\quad U^{\prime }(x_{i}^{\ast })=\xi _{i}.
\label{key}
\end{equation}
\end{pr}

The proof of Proposition \ref{PR0} given in Appendix \ref{A0} uses the
concavity of $U$ and is a basic application of pathwise optimization. It is
clear from Proposition \ref{PR0} that the utility function $U(\cdot )$ that
rationalizes the optimal investment choice $X^{\ast }$ in this setting is
not unique. The candidate utility $U_{P}(\cdot )$ proposed by Peleg and
Yaari \citeyear{PY75} is given by 
\begin{equation*}
U_{P}(x)=\int_{0}^{x}v(y)dy
\end{equation*}%
where $v(y)=\xi _{1}-y+x_{1}^{\ast }$ for $y<x_{1}^{\ast }$ and $v(y)=\xi
_{j}+(y-x_{j}^{\ast })\frac{(\xi _{j+1}-\xi _{j})}{(x_{j+1}^{\ast
}-x_{j}^{\ast })}$ for $y\in \lbrack x_{j}^{\ast },x_{j+1}^{\ast })$ $%
(j=1,2,...,N-1)$ and $v(y)=\xi _{N}$ for $y\geq x_{N}^{\ast }$. $U_{P}(\cdot
)$ is differentiable at all $x_{i}^{\ast }$ and satisfies the condition $%
U_{P}^{\prime }(x_{i}^{\ast })=\xi _{i}$ of Proposition \ref{PR0}. It will
be clear later that if we use this utility and generalize the financial
market to have a continuum of states, then $X^{\ast }$ no longer solves %
\eqref{RE2}. In this paper, we introduce\footnote{%
The formula for $U(\cdot )$ in \eqref{UTI} will appear more intuitive after
reading the proof of Theorem \ref{th1} in the following section. It is built
so that the optimum for the expected utility problem \eqref{RE2} corresponds
to the cheapest strategy with distribution $F$.} a utility function $U(\cdot
)$ 
\begin{equation}
U(x)=\int_{x_{1}^{\ast }}^{x}F_{\xi }^{-1}(1-F(y))dy,  \label{UTI}
\end{equation}%
where $F(\cdot )$ is the distribution function (cdf) of $X^{\ast }$, $F_{\xi
}(\cdot )$ is the cdf of $\xi $ and the quantile function $F_{\xi }^{-1}$ is
defined as $F_{\xi }^{-1}(p)=\inf \{t\ |\ F_{\xi }(t)\geq p\}$ with the
convention that $F_{\xi }^{-1}(0)=0$. One observes that $y\mapsto F_{\xi
}^{-1}(1-F(y))$ is a decreasing right-continuous step function (taking the
value $\xi _{1}$ for $y<x_{1}^{\ast }$, $\xi _{i+1}$ when $y\in \lbrack
x_{i}^{\ast },x_{i+1}^{\ast })$ for $i=1,...,N-1$, and the value $0$ 
%
for $y\geq x_{N}^{\ast }$). Thus, $U(\cdot )$ is continuous, piecewise
linear and concave satisfying $U(x_{1}^{\ast })=0$. It is differentiable at
all points except at each $x_{i}^{\ast }$, $i=1,2,...,N$, where it has a
distinct right and left derivative. Denote by $U^{\prime }(\cdot )$ the left
derivative of $U(\cdot )$. We have that condition \eqref{key} of Proposition %
\ref{PR0} is satisfied. Moreover, Theorem \ref{th3} will show that it is the
unique (generalized) utility that explains the demand for $X^{\ast }$ in the
market when $\xi $ is continuously distributed.

Henceforth, we assume that the state price $\xi $ is continuously
distributed, and we extend the above construction to the more general market
(infinite state space) in Section \ref{S1}. Doing so is not only a technical
extension, but rather natural in the context of making optimal investment
choices and at least consistent with the literature on it. As the example
illustrates, it is also necessary in order to define \textit{the} utility
function (up to a linear transformation) that rationalizes the demand for a
given optimal consumption $X^{\ast },$ and that can thus be used to compute
the implied risk aversion of the investor.

When requiring that the utility function is differentiable at all points, we
will show that we can only explain continuous distributions. 
However, there are many situations in which the investor wants a discrete
distribution of wealth or a mixed distribution. He and Zhou \citeyear{HZ}
show that, under some assumptions, optimal payoffs in Yaari's dual theory
have a discrete distribution, whereas in the case of cumulative prospect
theory, the optimal final wealth has a mixed distribution. While these
observations point to differences between the decision theories, we show
that these optimal payoffs can be rationalized by (generalized) expected
utility theory. Section \ref{S2} provides some applications of the results
derived in Section \ref{S1}. In particular, we illustrate how a
non-decreasing concave utility function can be constructed to explain the
demand for optimal investment in Yaari's \citeyear{yaari} setting. 
One of the key findings presented toward the end of the paper (Section \ref%
{S3}) is to infer risk aversion and to show that DARA is equivalent to a
demand for terminal wealth that exhibits more spread than the market
variable $H_{T}:=-\log (\xi _{T})$. 

\section{Setting \label{Sec:Setting}}

We assume an arbitrage-free and frictionless financial market $(\Omega ,{%
\mathcal{F}},\mathbb{P})$ with a fixed investment horizon of $T>0$. Let $\xi
_{T}$ be the pricing kernel that is agreed upon by all agents. We assume
that it has a positive density on $\mathbb{R}^{+}\backslash \{0\}$. The
value $X_0$ at time $0$ of a consumption $X_{T}$ at $T$ is then computed as 
\begin{equation*}
X_0=E[\xi _{T}X_{T}].
\end{equation*}%
We consider only terminal consumptions $X_{T}$ such that $X_0$ is finite.
Throughout the paper, agents have law-invariant and non-decreasing
preferences $V(\cdot )$. 
Theorem \ref{LI} shows that these properties are equivalent to preferences $%
V(\cdot )$ that respect FSD. $X\prec _{fsd}Y$ means that $Y$ is
(first-order) stochastically larger than $X$, i.e., for all $x\in \mathbb{R} 
$, $F_{X}(x)\geq F_{Y}(x),$ where $F_{X}$ and $F_{Y}$ denote the cumulative
distribution functions (cdfs) of $X$ and $Y$ respectively. Equivalently for
all non-decreasing functions $v$, $E[v(X)]\le E[v(Y)]$.

\begin{theo}
\label{LI} Preferences $V(\cdot )$ are non-decreasing and law-invariant if
and only if $V(\cdot )$ satisfies FSD.
\end{theo}

\noindent The proof of Theorem \ref{LI} can be found in Appendix \ref{appLI}%
. An agent with preferences $V(\cdot )$ finds her optimal terminal
consumption $X_{T}^{\ast }$ by solving the following optimization problem: 
\begin{equation}
\max_{X_{T}\ |\ E[\xi _{T}X_{T}]=X_0}V(X_{T}).  \label{EUopt}
\end{equation}

When an optimal strategy $X_{T}^{\ast }$ exists, denote its cdf by $F^\ast$.
Intuitively, since $V(\cdot )$ is non-decreasing and law-invariant, then
among all strategies with cdf $F$, the optimum $X_{T}^{\ast }$ must be the
cheapest possible one. This observation is made precise in the following
lemma, which is instrumental to the rest of the paper. We omit the proof, as
it is proved in Bernard, Boyle and Vanduffel \citeyear{BBV}.\footnote{%
The first part of the lemma corresponds to their Proposition 5. The second
part corresponds to their Proposition 2 and Corollary 2. It is also closely
related to results that first appeared in Dybvig %
\citeyear{dybvigJoB,dybvigRFS}.}

\begin{lem}[Cost-efficiency]
\label{th0} Assume that an optimum $X_{T}^{\ast }$ of \eqref{EUopt} exists
and denote its cdf by $F$. Then, $X_{T}^{\ast }$ is the cheapest
(cost-efficient) way to achieve the distribution $F$ at the investment
horizon $T$, i.e., $X_{T}^{\ast }$ also solves the following problem: 
\begin{equation}
\min_{X_{T}|X_{T}\sim F}E[\xi _{T}X_{T}].  \label{CEb}
\end{equation}%
Furthermore, for any given cdf $F$, the solution $X_{T}^{\star }$ to Problem %
\eqref{CEb} is almost surely (a.s.) unique and writes as $X_{T}^{\star
}=F^{-1}(1-F_{\xi _{T}}(\xi _{T}))$. Payoffs are cost-efficient if and only
if they are non-increasing in the pricing kernel $\xi _{T}.$
\end{lem}

Lemma \ref{th0} provides us with an alternative approach to portfolio
optimization. Usually, one resorts to a value function $V(\cdot )$ in order
to model preferences and then finds the optimal consumption by solving
Problem \eqref{EUopt}. Using Lemma \ref{th0}, one specifies a desired
distribution $F $ of terminal wealth up-front\footnote{%
As aforementioned, it is presumably easier for many investors to describe a
target terminal wealth distribution $F$ than to articulate the value
function $V(\cdot )$ governing their investment decision (see e.g.,
Goldstein, Johnson and Sharpe {\citeyear {GJS2}}).} and determines the
cheapest strategy that is distributed with $F$. 

In general, there may be more than one solution to \eqref{EUopt}. However,
two different solutions must have different cdfs because the cost-efficient
payoff generating a given distribution is unique. In the context of EUT, $%
V(X_{T})=E$($U(X_{T}))$ for some utility function $U(\cdot ).$ When $U(\cdot
)$ is not concave, a standard approach to solving Problem \eqref{EUopt} is
to introduce the concave envelope of $U(\cdot )$, denoted by $U_{C}(\cdot )$%
, which is the smallest concave function larger than or equal to $U(\cdot )$%
. Reichlin \citeyear{Reichlin2012} shows that under some technical
assumptions, the maximizer for $U_{C}(\cdot )$ is also the maximizer for $%
U(\cdot )$. However, this maximizer is only unique under certain cases (see
Lemma 5.9 of Reichlin \citeyear{Reichlin2012}).

%
In the following section, we reconcile different decision theories by
showing that an optimal portfolio in any behavioral theory that respects FSD
can be obtained as an optimal portfolio for a risk averse investor
maximizing a (generalized) expected utility.


\section{Explaining Distributions through Expected Utility Theory \label{S1}}

In the first part of this section, we show how the traditional expected
utility setting with a strictly increasing and strictly concave utility
function on an interval can be used to explain $F$ when it is strictly
increasing and continuous on this interval. The second part shows that, more
generally, any distribution of optimal final wealth can be explained using a
\textquotedblleft generalized\textquotedblright\ utility function defined
over $\mathbb{R}$. Since a solution $X_{T}^{\ast }$ to the optimization
problem \eqref{EUopt} is completely characterized by its distribution $F$,
it follows that $X_{T}^{\ast }$ is also optimal for some expected utility
maximizer. The last part discusses tests for verifying whether investor
behavior can be rationalized by the (generalized) Expected Utility Theory.

\subsection{Standard Expected Utility Maximization}


\begin{defi}[${\mathcal{U}}_{(a,b)}$: set of utility functions]
\label{DEFIStandard} Let $(a,b)\subset \mathbb{R}$ where $a,b\in \overline{%
\mathbb{R}}$ (with $\overline{\mathbb{R}}=\mathbb{R}\cup \{-\infty ,+\infty
\}$). We define a set $\mathcal{U}_{(a,b)}$ of utility functions $U$ on $%
(a,b)$ such that $U:(a,b)\rightarrow \mathbb{R}$ is continuously
differentiable, strictly increasing on $(a,b)$, $U^{\prime }$ is strictly
decreasing on $(a,b)$ (so that the investor is risk averse), $U(c)=0$ for
some $c\in (a,b)$, $U^{\prime }(a):=\lim_{x\searrow a}U^{\prime }(x)=+\infty 
$, and $U^{\prime }(b):=\lim_{x\nearrow b}U^{\prime }(x)=0$.
\end{defi}

Note that Inada's conditions correspond to $a=c=0$ and $b=+\infty$.

\begin{defi}[Rationalization by Standard Expected Utility Theory]
An optimal portfolio choice $X^*_T$ with a finite budget $X_0$ is
rationalizable by the standard expected utility theory if there exists a
utility function $U\in{\mathcal{U}}_{(a,b)}$ such that $X_T^*$ is also the
optimal solution to 
\begin{equation}
\max_{X\ |\ E[\xi X]=X_{0}}E[U(X)].  \label{EUpb}
\end{equation}
\end{defi}

The following lemma finds the optimal payoff for an expected utility
maximizer with a utility function in $\mathcal{U}_{(a,b)}$.

\begin{lem}
\label{lemma1}Consider a utility function $U$ in $\mathcal{U}_{(a,b)}$.
Assume that $X_{0}\in (E[\xi _{T}a],E[\xi _{T}b]).$ The (a.s.) unique
optimal solution $X_{T}^{\star }$ to the expected utility maximization %
\eqref{EUpb} is given by 
\begin{equation*}
X_{T}^{\star }:=\left[ U^{\prime }\right] ^{-1}\left( \lambda ^{\ast }\xi
_{T}\right) 
\end{equation*}%
where $\lambda ^{\ast }$ $>0$ is such that $E\left[ \xi _{T}X_{T}^{\ast }%
\right] =X_{0}$. Furthermore, $X_{T}^{\star }$ has a continuous distribution 
$F$, which is strictly increasing on $(a,b)$ with $F(a^{+})=0,$ $F(b^{-})=1$.
\end{lem}

This lemma is proved by Merton \citeyear{M} and by Cox and Huang %
\citeyear{CH} when Inada's conditions are satisfied. Note in particular that
in the statement of the lemma, the condition on the budget automatically
disappears when Inada's conditions are satisfied. The result is presented
here in a slightly more general setting, as it will be needed in what
follows. The proof of this lemma (given in Appendix \ref{Lem1proof})
illustrates how the pathwise optimization technique (used repeatedly
throughout the paper) can help to solve the standard expected utility
maximization \eqref{EUpb}. The following theorem 
gives, for any strictly increasing continuous distribution of final wealth,
an explicit construction of the utility function that explains the
investor's demand in the expected utility maximization framework.

\begin{theo}[Strictly increasing continuous distribution]
\label{th1} Consider a strictly increasing and continuous cdf $F$ on $%
(a,b)\subset 
\mathbb{R}
$ with $a,b\in \overline{\mathbb{R}}$. Assume that the cost of the unique
cost-efficient payoff $X_{T}^{\star }$ solving \eqref{CEb} is finite and
denote it by $X_0$. Then $X_{T}^{\star }$ is also the optimal solution of
the expected utility maximization problem \eqref{EUpb} with the following
explicit utility function $U\in\mathcal{U}_{(a,b)}$ 
\begin{equation}
U(x)=\int_{c}^{x}F_{\xi _{T}}^{-1}(1-F(y))dy  \label{Utility1}
\end{equation}%
for some $c$ such that $F(c)>0$. This utility $U$ is unique in $\mathcal{U}%
_{(a,b)}$ up to a linear transformation.
\end{theo}

\proof\ When $U(\cdot)$ satisfies the conditions of Lemma \ref{lemma1}, the
optimal solution to \eqref{EUpb} can be written as 
\begin{equation}
X_{T}^{\star }:=\left[ U^{\prime }\right] ^{-1}\left( \lambda^\star \xi
_{T}\right) ,  \label{opt}
\end{equation}%
where $\lambda^* >0$ is chosen such that $E[\xi _{T}X_{T}^{\star }]=\omega
_{0} $. $\lambda^\star $ exists because of the conditions on $U$ (see also
Lemma \ref{lemma1}). From Lemma \ref{th0}, the unique cost-efficient payoff
with cdf $F$ has the following expression 
\begin{equation}
X_{T}^{\star }=F^{-1}\left( 1-F_{\xi _{T}}(\xi _{T})\right).  \label{optopt}
\end{equation}
The utility function $U(\cdot )$ as defined in \eqref{Utility1} is
constructed to equate \eqref{opt} and \eqref{optopt}. Specifically, using
the properties of continuity and increasingness of $F_{\xi_T}$, $U(\cdot )$
belongs to $\mathcal{U}_{(a,b)}$ and thus satisfies all conditions of Lemma %
\ref{lemma1}: it is continuously differentiable on $(a,b)$ and for $z\in
(a,b)$, $U^{\prime }(z):=F_{\xi _{T}}^{-1}\left( 1-F(z)\right)$. Then $%
U(\cdot )$ is strictly increasing and $U^{\prime }(\cdot )$ is strictly
decreasing on $(a,b)$. Note also that $U(c)=0$, the limit of $U^\prime(x)$
is $+\infty$ when $x\searrow a$ and the limit of $U^\prime(x)$ is 0 when $%
x\nearrow b$ because of $F_{\xi_T}^{-1}(1)=+\infty$ and $F_{\xi_T}^{-1}(0)=0$%
. In addition, $U(\cdot )$ is such that 
\begin{equation*}
\left[ U^{\prime }\right] ^{-1}\left( \lambda ^{\star }\xi _{T}\right)
=F^{-1}\left( 1-F_{\xi _{T}}(\lambda ^{\star }\xi _{T})\right) .
\end{equation*}%
%
%
%
%
%
%
%
%
%
%
%
%
%
%
%
%
%
%
%
%
%
%
%
%
%

Observe that for $\lambda^\star =1,$ $X_{T}^{\star }\sim F$. 
By assumption, $X_0$ is the cost of the unique cost-efficient payoff with
distribution $F$. Therefore, $\lambda^\star=1$ ensures that $E[\xi
_{T}X_{T}^{\star }]=X_0$ and $X_{T}^{\star }=\left[ U^{\prime}\right]
^{-1}\left(\xi _{T}\right)$ is thus the solution to the maximum expected
utility problem \eqref{EUpb}. If there is another utility $U_2\in\mathcal{U}%
_{(a,b)}$ such that $X_T^*$ is an optimal solution to \eqref{EUpb}, then, by
the same reasoning as above, we find that $\xi_T=U_2^\prime(X^*_T)=U^%
\prime(X^*_T)$ a.s.. Since $X^*_T$ has a strictly increasing distribution on 
$(a,b)$ (with a positive density everywhere on $(a,b)$), $%
U_2^\prime(x)=U^\prime(x)$ for all $x\in(a,b)$, and thus $U_2$ can be
obtained in terms of a linear transformation of $U$. 
\hfill $\Box $

If an expected utility maximizer chooses a particular investment with
distribution of terminal wealth $F$, then the only utility function in $%
\mathcal{U}_{(a,b)}$ that rationalizes her choice is given by %
\eqref{Utility1} (up to a linear transformation). Note also that this
utility function \eqref{Utility1} involves properties of the financial
market at the horizon time $T$ (through the cdf $F_{\xi_T}$ of the pricing
kernel $\xi _{T}$).

In the second part of this section, we generalize Theorem \ref{th1} to
include more general distributions (discrete and mixed distributions).
Obviously, any 
distribution $F$ can always be approximated by a sequence of continuous
strictly increasing distributions, $F_{n}$. Then, for each $F_{n},$ Theorem %
\ref{th1} allows us to obtain the corresponding strictly concave and
strictly increasing utility function $U_{n}\in\mathcal{U}_{(a,b)}$ so that
the optimal investment for an expected utility maximizer with utility
function $U_{n}$ is distributed with the cdf $F_{n}$. Thus, Theorem \ref{th1}
already explains \emph{approximately} the demand for all distributions.


\paragraph{Rationalization by Standard Expected Utility Theory\newline
}

To summarize our findings in this section, we formulate the following
characterization of consumptions that can be rationalizable by standard
expected utility theory.

\begin{theo}[Rationalizable consumption by Standard EUT]
\label{thD} Consider a terminal consumption $X_{T}$ at time $T$ purchased
with an initial budget $X_{0}$ and distributed with a \textit{continuous}
cdf $F$. The eight following conditions are equivalent:\vspace{-3mm}

\begin{itemize}
\item[(i)] $X_{T}$ is rationalizable by standard Expected Utility Theory.

\item[(ii)] $X_T$ is cost-efficient with cdf $F$.

\item[(iii)] $X_{0}=E[\xi _{T}F^{-1}(1-F_{\xi _{T}}(\xi _{T}))]$.

\item[(iv)] $X_{T}=F^{-1}(1-F_{\xi _{T}}(\xi _{T}))$ a.s.

\item[(v)] $X_{T}$ is non-increasing in $\xi _{T}$ a.s.

\item[(vi)] $X_{T}$ is the solution to a maximum portfolio problem for some
objective $V(\cdot )$ that satisfies FSD.

\item[(vii)] $X_{T}$ is the solution to a maximum portfolio problem for some
law-invariant and non-decreasing objective function $V(\cdot )$.

\item[(viii)] $X_{T}$ is the solution to a maximum portfolio problem for
some objective $V(\cdot )$ that satisfies SSD.
\end{itemize}
\end{theo}

Note that Theorem \ref{thD} highlights the strong link between the concept
of \textquotedblleft cost-efficiency\textquotedblright\ and
\textquotedblleft rationalization\textquotedblright\ of consumption by
Expected Utility Theory. Recall that $X\prec _{fsd}Y$ means that for all
non-decreasing utility functions $E[u(X)]\leq E[u(Y)],$ and $X\prec _{ssd}Y$
means that the expected utilities are ordered for concave and non-decreasing
utility functions. It is thus clear that if $V$ satisfies SSD, then it
satisfies FSD so that $(viii)$ implies $(vi)$ trivially. 
Our results show that the converse is true and thus that for the
rationalization of consumption by Expected Utility Theory it is only
required that preferences satisfy FSD, which, unlike the case of SSD, is an
assumption that is postulated by most common decision theories.

\begin{re}
Theorem \ref{thD} cannot be generalized in a discrete setting when the
states are not equiprobable as $(vi)$ is not equivalent to $(i)$.\footnote{%
Using Proposition \ref{PR0} and the results of \cite{BBV} in the discrete
setting, it is possible to prove a discrete version of Theorem \ref{thD} 
assuming states are equiprobable.} For example, take $\Omega =\{\omega
_{1},\omega _{2}\}$ with $P(\omega _{1})=\frac{1}{3}$ and $P(\omega _{2})=%
\frac{2}{3},$ and $\xi _{1}=\frac{3}{4}$ and $\xi _{2}=\frac{9}{8}.$ Take a
budget $X_{0}=1$ and consider $X$ with $X(\omega _{1})=a_{1}$ and $X(\omega
_{2})=a_{2}$ satisfying the budget condition $\frac{a_{1}}{4}+\frac{3a_{2}}{4%
}=1.$ Let the objective be defined as $V(X):=VaR_{1/3}^{+}(X)\mathds{1}%
_{P(X<0)=0}$ (where $VaR_{\alpha }^{+}(X)$ is defined as $VaR_{\alpha
}^{+}(X):=\sup \{x\in \mathbb{R},F_{X}(x)\leq \alpha \}$). Note that $%
V(\cdot )$ is clearly law-invariant and non-decreasing. Thus, $V(\cdot )$
satisfies FSD. It is clear that $V(\cdot )$ is maximized for $X^{\ast }$
defined through $X^{\ast }(\omega _{1})=0$ and $X^{\ast }(\omega _{2})=\frac{%
4}{3}$. By contrast, we can show that $X^{\ast }$ is never optimal for an
expected utility maximizer with non-decreasing concave utility $U$ on $[0,%
\frac{4}{3}]$ (range of consumption). To this end, assume without loss of
generality that $U(0)=0$ and $U(\frac{4}{3})=1.$ Consider $Y$ such that $%
Y(\omega _{1})=\frac{4}{3}$ and $Y(\omega _{2})=\frac{8}{9}$. Observe that $%
E[\xi Y]=E[\xi X^{\ast }]=1$. By concavity of $U,$ $\frac{U(\frac{8}{9})-U(0)%
}{8/9}\geq \frac{U(\frac{4}{3})-U(0)}{4/3}$, and thus $U(\frac{8}{9})\geq 
\frac{2}{3}.$ Hence, $E[U(Y)]=\frac{1}{3}U(\frac{4}{3})+\frac{2}{3}U(\frac{8%
}{9})\geq \frac{7}{9}>E[U(X^{\ast })]=\frac{2}{3}$.
\end{re}

The equivalence of $(vii)$ and $(viii)$ shows that Afriat's \citeyear{A}
theory shares a similar idea, as explained in the following remark.

\begin{re}
\label{rkk} Consider observations of prices and quantities for a given set
of goods over some period of time ($t=1,2,...,n$). At time $t$, one observes
a vector of prices ${\mathbf{p}_{t}}=(p_{t}^{1},...,p_{t}^{n})$ for $n$
goods as well as the respective quantities of each good purchased ${\mathbf{q%
}_{t}}=(q_{t}^{1},...,q_{t}^{n})$. Afriat \citeyear{A} aims at finding a
utility function $u$ such that at any time $t$, 
\begin{equation*}
u({\mathbf{q}_{t}})=\max_{{\mathbf{q}}}\left\{ u(\mathbf{q})\ |\ {\mathbf{p}%
_{t}^{t}}\cdot {\mathbf{q}}\leq {\mathbf{p}_{t}^{t}}\cdot {\mathbf{q}_{t}}%
\right\},
\end{equation*}%
where $u(\mathbf{q})$ is to be understood as the utility of consuming ${%
\mathbf{q}}$. Note that ${\mathbf{p}_{t}^{t}}\cdot {\mathbf{q}_{t}}$ is the
price of the consumption bundle ${\mathbf{q}_{t}}$ and the constraint ${%
\mathbf{p}_{t}^{t}}\cdot {\mathbf{q}}\leq {\mathbf{p}_{t}^{t}}\cdot {\mathbf{%
q}_{t}}$ reflects the fact that ${\mathbf{q}}$ was also an available
consumption at time $t$. Under some conditions (called cyclical
consistency), Afriat \citeyear{A} proves the existence of a strictly
increasing, continuous and concave function. In his recent review on
revealed preferences, Vermeulen \citeyear{vermeulen2012foundations} explains
that a \textit{\textquotedblleft violation of cyclical consistency implies
that a consumption bundle that is revealed} [as] \textit{preferred to some
other bundle is strictly less expensive than} [the] \textit{other bundle.
Nevertheless, the consumer has chosen the revealed worse and more expensive
bundle, which implies that this consumer violates the utility maximization
hypothesis''.} We observe that our condition for a terminal consumption
(portfolio) to be an optimum for an expected utility maximizer is also
related directly to its cost and to the fact that it is the cheapest
consumption with that distribution (cost-efficient consumption).
\end{re}

Theorem \ref{thD} can be used to test whether or not an observed demand for
a terminal consumption can be rationalized by expected utility theory. 
Assume that an investor chooses to invest his initial budget $X_{0}$ from $0$
to time $T$ in such a way that he is generating a payoff $X_{T}$ at time $T$%
. A simple test to investigate whether this investment choice is
rationalizable by Expected Utility Theory consists in checking conditions
(ii) or (iii) of the theorem. This idea corresponds to the steps taken by
Amin and Kat \citeyear{AK} to determine in another context whether a hedge
fund is efficient. \footnote{%
Specifically, they observe the initial investment in a hedge fund and
discuss whether the distribution of returns of the fund could have been
obtained in a cheaper way.} Note that this test is based only on the cost of
the payoff. If the distribution of returns is not obtained in the cheapest
way, it may be caused by an optimal investment criterion that does not
satisfy FSD. This is also a potential explanation for the pricing kernel
puzzle (Brown and Jackwerth \citeyear{brown2004pricing}, Hens and Reichlin %
\citeyear{hens2013three}).


\subsection{Generalized Expected Utility Maximization}

In the previous section, we used cost-efficiency to construct a utility
function that is continuously differentiable, strictly concave and strictly
increasing on an interval $(a,b)$ with $a,b\in \overline{\mathbb{R}}$ to
explain the demand for continuous and strictly increasing distributions on $%
(a,b)$. The same approach can be used to construct a \textquotedblleft
generalized\textquotedblright\ utility function defined on the entire real
line $\mathbb{R}$, which explains the demand for \textit{any} distribution.
A generalized utility function does not need to be either differentiable on $%
(a,b)$ or strictly concave. It is formally defined as follows.

\begin{defi}[$\protect\widetilde{\mathcal{U}}_{(a,b)}$: set of generalized
utility functions]
\label{DEFI} Let $(a,b)\subset \mathbb{R}$. We say that $\widetilde{U}:{%
\mathbb{R}}\rightarrow \overline{\mathbb{R}}$ belongs to the set of
generalized utility functions $\widetilde{\mathcal{U}}_{(a,b)}$ if $%
\widetilde{U}(x)$ writes as 
\begin{equation*}
\widetilde{U}(x):=\left\{ 
\begin{array}{ll}
U(x) & \hbox{for}\ x\in (a,b), \\ 
-\infty & \hbox{for}\ x<a, \\ 
U(a^{+}) & \hbox{for}\ x=a, \\ 
U(b^{-}) & \hbox{for}\ x\geq b,%
\end{array}%
\right.
\end{equation*}
where $U:(a,b)\rightarrow \mathbb{R} $ is strictly increasing and concave on 
$(a,b)$. We then define  $\widetilde{U}^{\prime }$ (with abuse of notation)
on $\mathbb{R}$ as follows. On $(a,b),$ $\widetilde{U}^{\prime }$ denotes
the left derivative of $\widetilde{U}.$ For $x<a,$ $\widetilde{U}^{\prime
}(x):=+\infty .$ $\widetilde{U}^{\prime }(a):=\lim_{x\searrow a}U^{\prime
}(x)$ and $\widetilde{U}^{\prime }(b):=\lim_{x\nearrow b}U^{\prime }(x),$
and for $x>b$, $\widetilde{U}^{\prime }(x)=0$. Conventions: if $a=-\infty $
then $\widetilde{U}(a):=\lim_{x\searrow a}U(x)$ and $\widetilde{U}^{\prime
}(a):=+\infty ,$ if $b=\infty $ then $\widetilde{U}(b):=\lim_{x\nearrow
b}U(x)$ and $\widetilde{U}^{\prime }(b):=0.$
\end{defi}


\begin{defi}[Rationalization by Generalized Expected Utility Theory]
An optimal portfolio choice $X^*_T$ with a finite budget $X_0$ is
rationalizable by the generalized standard expected utility theory if there
exists a utility function $\widetilde U\in\widetilde{\mathcal{U}}_{(a,b)}$
such that $X_T^*$ is also the optimal solution to 
\begin{equation}
\max_{X_{T}\ |\ E[\xi _{T}X_{T}]=X_0}E\left[ \widetilde{U}(X_{T})\right].
\label{eu2b}
\end{equation}
\end{defi}

The following lemma finds the optimal payoff for a generalized expected
utility maximizer with a utility function in $\widetilde{\mathcal{U}}%
_{(a,b)} $.

\begin{lem}
\label{lemma1bis}Consider a generalized utility function $\widetilde{U}$ of $%
\widetilde{\mathcal{U}}_{(a,b)}$. Assume that $X_{0}\in (E[\xi _{T}a],E[\xi
_{T}b]).$ The optimal solution $X_{T}^{\star }$ to the generalized expected
utility maximization \eqref{eu2b} exists, is a.s. unique and is given by 
\begin{equation*}
X_{T}^{\star }:=\left[ \widetilde{U}^{\prime }\right] ^{-1}\left( \lambda
^{\ast }\xi _{T}\right)
\end{equation*}%
where $\widetilde{U}^{\prime }$ is as defined 
in Definition \ref{DEFI} and where $\lambda ^{\ast }$ $>0$ is such that $E%
\left[ \xi _{T}X_{T}^{\star }\right] =X_{0}$ and 
\begin{equation}
\left[ {\tilde{U}}^{\prime }\right] ^{-1}(y):=\inf \left\{ x\in (a,b)\ |\ {%
\tilde{U}}^{\prime }(x)\leq y\right\} ,  \label{TU}
\end{equation}%
with the convention that $\inf \{\emptyset \}=b$. Furthermore, $X_{T}^{\star
}$ has potential mass points.
\end{lem}

Lemma \ref{lemma1bis} is proved in Appendix \ref{lemma1bisProof} and allows
us to derive a unique implied generalized expected utility to rationalize
the demand for any distribution.

\begin{theo}
\label{th3} Let $F$ be a distribution. Let $X_{T}^{\star }$ be the optimal
solution to \eqref{CEb} for the cdf $F$. Denote its cost by $X_0$ and assume
that it is finite: then $X_{T}^{\star }$ is also the optimal solution to
Problem \eqref{eu2b} where $\tilde{U}:$ $%
\mathbb{R}
\rightarrow \overline{\mathbb{R}}$ is a generalized utility function defined
as 
\begin{equation}
\tilde{U}(x):=\int_{c}^{x}F_{\xi _{T}}^{-1}(1-F(y))dy
\label{FormulaUtility2}
\end{equation}
with some $c\ge a$ such that $F(c)>0.$ 
Conventions: $F_{\xi_T}^{-1}(1)=+\infty$, $F_{\xi_T}^{-1}(0)=0$, if $x_1<x_2$
then $\int_{x_1}^{x_2}(+\infty)dy=+\infty$, and $\int_{x_2}^{x_1}(+%
\infty)dy=-\infty$ and $\int_{x_1}^{x_1}g(y)=0$ for all $g$ valued in $\bar{%
\mathbb{R}}$. $\tilde{U}(\cdot)$ is unique in the class of generalized
utilities up to a linear transformation.
\end{theo}

\paragraph{Rationalization by Generalized Expected Utility Theory\label%
{testi}\newline
}

We can summarize our findings in this section by the following theorem,
which is similar to Theorem \ref{thD} but now includes all possible
distributions of final wealth. 

\begin{theo}
\label{thD2} Consider a terminal consumption $X_{T}$ at time $T$ purchased
with an initial budget $X_{0}$ and distributed with $F$. The following eight
conditions are equivalent:\vspace{-3mm}

\begin{itemize}
\item[(i)] $X_{T}$ is rationalizable by Generalized Expected Utility Theory.

\item[(ii)] $X_T$ is cost-efficient.

\item[(iii)] $X_{0}=E[\xi _{T}F^{-1}(1-F_{\xi _{T}}(\xi _{T}))]$.

\item[(iv)] $X_{T}=F^{-1}(1-F_{\xi _{T}}(\xi _{T}))$ a.s.

\item[(v)] $X_{T}$ is non-increasing in $\xi _{T}$ a.s.

\item[(vi)] $X_{T}$ is the solution to a maximum portfolio problem for some
objective $V(\cdot )$ that satisfies FSD.

\item[(vii)] $X_{T}$ is the solution to a maximum portfolio problem for some
law-invariant and non-decreasing objective function $V(\cdot )$.

\item[(viii)] $X_{T}$ is the solution to a maximum portfolio problem for
some objective $V(\cdot )$ that satisfies SSD.
\end{itemize}
\end{theo}

The proof of this theorem is identical to that of Theorem \ref{thD} by
replacing expected utility with generalized expected utility and is thus
omitted.

\section{From Distributions to Utility Functions \label{S2}}

In this section, we use the results of the previous sections to derive
utility functions that explain the demand for some financial products and
for some distributions of final wealth for agents with preferences
satisfying FSD. Let us start with a simple example showing how one can
recover the popular CRRA utility function from a lognormally distributed
final wealth. This example is particularly useful when explaining the
optimal demand for a retail investor who chooses an equity-linked structured
product with capital guarantee. The last example deals with an example of
non-expected utility: Yaari's dual theory of choice. In this case, we are
able to exhibit the non-decreasing concave utility function such that the
optimal strategy in Yaari's \citeyear{yaari} theory is also obtained in an
expected utility maximization framework.%

For ease of exposition, we restrict ourselves to the one-dimensional
Black-Scholes model with one risky asset, $S_{T}$.\footnote{%
All developments can be executed in the general market setting given in
Section \ref{Sec:Setting}. However, closed-form solutions are more
complicated or unavailable.} In this case, the pricing kernel $\xi _{T}$ is
unique and can be expressed explicitly in terms of the stock price $S_{T}$
as follows: 
\begin{equation}
\xi _{T}=\alpha \left( \frac{S_{T}}{S_{0}}\right) ^{-\beta },
\label{xiTform}
\end{equation}%
where $\frac{S_{T}}{S_{0}}\thicksim \mathcal{LN}\left( (\mu -\frac{\sigma
^{2}}{2})T,\text{ }\sigma ^{2}T\right) ,$ $\alpha =\exp \left( \frac{\theta 
}{\sigma }\left( \mu -\frac{\sigma ^{2}}{2}\right) T-\left( r+\frac{\theta
^{2}}{2}\right) T\right) $, $\beta =\frac{\theta }{\sigma }$ and $\theta =%
\frac{\mu -r}{\sigma }$. 
Formula (\ref{xiTform}) is well-known and can be found for example in
Section 3.3 of Bernard et al. \citeyear{BBV}. It follows that 
\begin{equation}
\xi _{T}\thicksim \mathcal{LN}\left( -rT-\frac{\theta ^{2}T}{2},\theta
^{2}T\right) .  \label{xsi}
\end{equation}

Assume first that consumption is restricted on $(0,\infty )$ and that the
investor wants to achieve a lognormal distribution $\mathcal{LN}(M,\Sigma^2)$
at maturity $T$ of her investment. The desired cdf is $F(x)=\Phi \left( 
\frac{\ln x-M}{\Sigma}\right) $, and from \eqref{xiTform} it follows that $%
F_{\xi _{T}}^{-1}(y)=\exp \left\{ \Phi ^{-1}(y)\theta \sqrt{T}-rT-\frac{%
\theta ^{2}T}{2}\right\} $. Applying Theorem \ref{th1}, the utility function
explaining this distribution writes as 
\begin{equation}
U(x)=\left\{ 
\begin{array}{ll}
a\frac{x^{1-\frac{\theta \sqrt{T}}{\Sigma}}}{1-\frac{\theta \sqrt{T}}{\Sigma}%
} & \frac{\theta \sqrt{T}}{\Sigma}\neq 1 \\ 
a\log (x) & \frac{\theta \sqrt{T}}{\Sigma}=1%
\end{array}%
\right. ,  \label{Lognormalutility}
\end{equation}%
where $a=\exp (\frac{M\theta \sqrt{T}}{\Sigma}-rT-\frac{\theta ^{2}T}{2}).$
This is a CRRA utility function with relative risk aversion $\frac{\theta 
\sqrt{T}}{\Sigma }$. A more thorough treatment of risk aversion is provided
in Section \ref{S3}.

\subsection{Explaining the Demand for Capital Guarantee Products}

Many structured products include a capital guarantee and have a payoff of
the form $Y_{T}=\max (G,S_{T}), $ where $S_{T}$ is the stock price and $K$
is the (deterministic) guaranteed level. $S_T$ has a lognormal distribution, 
${S_{T}}\sim \mathcal{LN}(M,\sigma^2T)$, where $M:=\ln S_0 + \left(\mu-\frac{%
\sigma^2}{2}\right)T$ so that the cdf for $Y_{T} $ is equal to $%
F_{Y_{T}}(y)= \mathds{1}_{y\geq G}\Phi \left( \frac{\ln y-M}{\sigma\sqrt{T}}%
\right)$. Since $Y_{T}$ has a mixed distribution (with mass point at $K$),
we can apply Theorem \ref{th3} to derive the corresponding utility function.
Let $p:=\Phi \left( \frac{\ln G-M}{\sigma\sqrt{T}}\right) $ and define the
following discrete and continuous cdfs: 
\begin{align}
F_{Y_{T}}^{D}(y)& =\left\{ 
\begin{array}{ll}
0 & y<G \\ 
1 & y\geq G%
\end{array}%
\right. , \quad F_{Y_{T}}^{C}(y) =\left\{ 
\begin{array}{ll}
0 & y<G \\ 
\frac{\Phi \left( \frac{\ln y-M}{\sigma\sqrt{T}}\right) -p}{1-p} & y\geq G%
\end{array}%
\right. .  \label{GteeFDY}
\end{align}%
Then, we can see that $F_{Y_{T}}(y)=pF_{Y}^{D}(y)+(1-p)F_{Y_{T}}^{C}(y)$. 
When $x<K$, $\tilde{U}(x)=-\infty $. 
The utility function $\widetilde U$ belongs to $\widetilde{\mathcal{U}}%
_{(K,\infty)}$ and is given by 
\begin{equation}
\tilde{U}(x)=\left\{ 
\begin{array}{ll}
-\infty & x<G, \\ 
a\frac{x^{1-\frac{\theta}{\sigma }}-G^{1-\frac{\theta}{\sigma }}}{1-\frac{%
\theta}{\sigma }} & x\geq G,\text{ }\frac{\theta }{\sigma }\neq 1, \\ 
a\log (\frac{x}{G}) & x\geq G,\text{ }\frac{\theta}{\sigma }=1,%
\end{array}%
\right.  \label{GteeUtil}
\end{equation}%
with $a=\exp (\frac{M\theta}{\sigma}-rT-\frac{\theta ^{2}T}{2}).$ The mass
point is explained by a utility that is infinitely negative for any level of
wealth below the guaranteed level. The CRRA utility above this guaranteed
level ensures the optimality of a lognormal distribution above the
guarantee, as aforementioned.

\subsection{Yaari's Dual Theory of Choice Model}

Optimal portfolio selection under Yaari's \citeyear{yaari} dual theory
involves maximizing the expected value of the terminal payoff under a
distorted probability function. Specifically, under Yaari's dual theory of
choice, decision makers evaluate the \textquotedblleft
utility\textquotedblright\ of their non-negative final wealth $X_{T}$ (with
cdf $F$) by calculating its distorted expectation $\mathds{H}_{w}\left[ X_{T}%
\right]: $ 
\begin{equation}
\mathds{H}_{w}\left[ X_{T}\right] =\int_{0}^{\infty }w\left( 1-{F}(x)\right) 
\mathrm{d}x\text{,}  \label{d1}
\end{equation}%
where the (distortion) function $w:\left[ 0,1\right] \rightarrow \left[ 0,1%
\right] $ is non-decreasing with $w(0)=0$ and $w(1)=1$.

The investor's initial endowment is $X_0\geq 0$. He and Zhou \citeyear{HZ}
find the optimal payoff when the distortion function is given by $%
w(z)=z^{\gamma }$ where $\gamma >1$. They show that there exists\footnote{%
It is proved that $c$ is the unique root on $\left( 1,\gamma e^{-rT}\right) $
of the function $\left[ \Phi \left( \frac{\ln x+rT+\frac{\theta ^{2}T}{2}}{%
\theta \sqrt{T}}\right) \right] ^{\gamma -1}\times\left[ x\Phi \left( \frac{%
\ln x+rT+\frac{\theta ^{2}T}{2}}{\theta \sqrt{T}}\right) -\gamma e^{-rT}\Phi
\left( \frac{\ln x+rT+\frac{\theta ^{2}T}{2}}{\theta \sqrt{T}}\right) \right]
$.} $c$ such that the optimal payoff is equal to 
\begin{equation}
X_{T}^{\star }=B\mathds{1}_{\xi _{T}\leq c},  \label{Dualpayoff}
\end{equation}%
where $B=X_0e^{rT}\Phi \left( \frac{\ln c+rT-\frac{\theta ^{2}T}{2}}{\theta 
\sqrt{T}}\right) >0$ is chosen such that the budget constraint $E[\xi
_{T}X_{T}^{\star }]=X_0$ is fulfilled. The corresponding cdf is 
\begin{equation}
F(x)=\left\{ 
\begin{array}{ll}
\Phi \left( \frac{-rT-\frac{\theta ^{2}T}{2}-\ln c}{\theta \sqrt{T}}\right)
& 0\leq x<B \\ 
1 & x\geq B%
\end{array}%
\right. .  \label{Dualcdf}
\end{equation}%
%
We find that the utility function $\widetilde{U}\in{\widetilde{\mathcal{U}}%
_{(0,B)}}$ is given by 
\begin{equation}
\tilde{U}(x)=\left\{ 
\begin{array}{ll}
-\infty & x<0, \\ 
c(x-c) & 0\leq x\leq B, \\ 
c(B-c) & x>B.%
\end{array}%
\right.  \label{DualUtil}
\end{equation}%
%
%
%
%
%
%
%
%
%
%
%
%
%
The utility function such that the optimal investment in the expected
utility setting is similar to the optimum in Yaari's framework %
\citeyear{yaari} is simply linear up to a maximum $c(B-c)$ and then constant
thereafter.

\section{From Distributions to Risk Aversion \label{S3}}


In this section, we compute risk aversion directly from the choice of the
distribution desired by the investor and from the distribution of the log
pricing kernel. We then show that it coincides with the Arrow-Pratt measures
for risk aversion when the distribution is continuous. 
Next, we show that decreasing absolute risk aversion (DARA) is equivalent to
terminal wealth exhibiting a spread greater than the market variable $%
H_{T}:=-\log (\xi _{T})$. In a Black-Scholes setting, agents thus have DARA
preferences if and only if they show a demand for distributions with tails
that are \textquotedblleft fatter than normal\textquotedblright . Our
characterization for DARA also allows us to construct an empirical test for
DARA preferences that is based on observed investor behavior.

For ease of exposition, we assume that all distributions in this section are
twice differentiable. We denote by $G$ the cdf of $H_{T}$ and by $g$ its
density, and we assume that $g(x)>0$ for all $x\in\mathbb{R}$. 

\subsection{Risk Aversion Coefficient}

\begin{defi}[Distributional Risk Aversion Coefficient]
Let $F$ be the distribution desired by the investor. Let $G$ and $g$ be the
cdf and density of $-\log(\xi_T)$, respectively. Consider a level of wealth $%
x$ such that $x=F^{-1}(p) $ for some $0<p<1$. We define the absolute and
relative risk aversion at $x$ as 
\begin{equation}
{\mathcal{A}}(x)=\frac{f(F^{-1}(p))}{g(G^{-1}(p))},\quad {\mathcal{R}}%
(x)=F^{-1}(p)\frac{f(F^{-1}(p))}{g(G^{-1}(p))}  \label{Absriskaversiondef}
\end{equation}
where 
\begin{equation}  \label{fdef}
f(y):=\lim_{\varepsilon\rightarrow0}\frac{F(y+\varepsilon)-F(y)}{\varepsilon}
\end{equation}
when this limit exists. If $F$ is differentiable at $y$, then $%
f(y)=F^{\prime}(y)$. If $F$ has a density, then $F^{\prime}(y)=f(y)$ at all $%
y$.
\end{defi}

Recall that the Arrow-Pratt measures for absolute and relative risk
aversion, ${\mathcal{A}}(x)$ resp. ${\mathcal{R}}(x)$, can be computed from
a twice differentiable utility function $U$ as ${\mathcal{A}}(x)=-\frac{%
U^{\prime \prime }(x)}{U^{\prime }(x)}$ and ${\mathcal{R}}(x)=x{\mathcal{A}}%
(x)$ .

\begin{theo}[Arrow-Pratt measures for risk aversion]
\label{RAVERSION} Consider an investor who targets a cdf $F$ for his
terminal wealth (with corresponding $f$ defined as \eqref{fdef}). Consider a
level of wealth $x$ in the interior of the support of the distribution $F$.
Then $x=F^{-1}(p) $ for some $0<p<1$. The Arrow-Pratt measures for absolute
and relative risk aversion at $x$ are given, respectively, as 
\begin{equation}
{\mathcal{A}}(x)=\frac{f(F^{-1}(p))}{g(G^{-1}(p))},\quad {\mathcal{R}}%
(x)=F^{-1}(p)\frac{f(F^{-1}(p))}{g(G^{-1}(p))}.  \label{Absriskaversion}
\end{equation}
\end{theo}

The proof of Theorem \ref{RAVERSION} is given in Appendix \ref{appRAVERSION}%
. Expression \eqref{Absriskaversion} shows that the coefficient for absolute
risk aversion can be interpreted as a likelihood ratio and is linked
directly to the financial market (through the cdf $G$ of the negative of the
log pricing kernel $H_T$).

\subsection{Decreasing Absolute Risk Aversion}

We provide precise characterizations of DARA in terms of distributional
properties of the final wealth and of the financial market. In what follows,
we only consider distributions that are twice differentiable.

\begin{theo}[Distributional characterization of DARA]
\label{SUFF} Consider an investor who targets some distribution $F$ for his
terminal wealth. The investor has (strictly) decreasing absolute risk
aversion (DARA) if and only if 
\begin{equation}
y\mapsto F^{-1}(G(y))\text{ is strictly convex on } \mathbb{R}.
\label{DARACondition-alt}
\end{equation}%
The investor has asymptotic DARA (DARA for a sufficiently high level of
wealth) if and only if there exists $y^{\star } \in \mathbb{R}$ such that $%
y\mapsto F^{-1}(G(y))$ is strictly convex on $(y^{\star },\infty )$.
\end{theo}

The proof of Theorem \ref{SUFF} is given in Appendix \ref{SUFFapp}. 
The convexity of the function $F^{-1}(G(x))$ reflects the fact that the
target distribution $F$ is \textquotedblleft
fatter-tailed\textquotedblright\ than the distribution $G$. In other words, $%
F$ is larger than $G$ in the sense of transform convex order (Shaked and
Shantikumar \citeyear{ShSh}, p. 214). 

Theorem \ref{SUFF} extends recent results by Dybvig and Wang \citeyear{DW} 
in another direction. These authors show that if agent $A$ has lower risk
aversion than agent $B, $ then agent $A$ purchases a distribution that is
larger than the other in the sense of SSD. Here, we show that the risk
aversion of an agent is decreasing in available wealth if and only if the
agent purchases a payoff that is heavier-tailed than the market variable $%
H_{T}$.

\begin{theo}
\label{SUFF2} Consider an investor with optimal terminal wealth $W_{T}\sim F$%
. The investor has decreasing absolute risk aversion if and only if $W_{T}$
is increasing and strictly convex in $H_{T}$.
\end{theo}

The proof of Theorem \ref{SUFF2} is given in Appendix \ref{SUFF2app}. Note
that an investor with optimal terminal wealth $W_{T}\sim F$ such that $F$
has right-bounded support does not exhibit DARA.\footnote{%
Recall that $k(x):=F^{-1}(G(x))$ is non-decreasing. Since $F$ is
right-bounded, there exists $b\in 
\mathbb{R}
,$ $k(x)\leq b$\ $(x\in 
\mathbb{R}
)$. We need to show that $k(x)$ cannot be strictly convex $(x\in 
\mathbb{R}
)$. We proceed by contradiction, so let $k(x)$ be strictly convex. This
implies that it remains above its tangent. As it is non-constant and
non-decreasing, there exists a point $x$ for which the tangent has a
positive slope and thus goes to infinity at infinity. It is thus impossible
for $k$ to be bounded from above. 
}


\subsection{The case of a Black-Scholes market}

In a Black-Scholes market we find from (\ref{xsi}) that $H_{T}$ is normally
distributed, with mean $rT+\frac{\theta ^{2}T}{2}$ and variance $\theta
^{2}T $. It is then straightforward to compute 
\begin{eqnarray*}
{\mathcal{A}}(x) =\theta f(x)\sqrt{2\pi T}\exp \left( \frac{1}{2}\left[ \Phi
^{-1}(1-F(x))\right] ^{2}\right), \quad {\mathcal{R}}(x) =\theta xf(x)\sqrt{%
2\pi T}\exp \left( \frac{1}{2}\left[ \Phi ^{-1}(1-F(x))\right] ^{2}\right).
\end{eqnarray*}
The financial market thus influences the risk aversion coefficient ${%
\mathcal{A}}(x)$ through the instantaneous Sharpe ratio $\theta =\frac{\mu -r%
}{\sigma }$. Interestingly, the effect of the financial market on the risk
aversion coefficient ${\mathcal{A}}(x)$ is proportional and does not depend
on available wealth $x $. This also implies that in a Black-Scholes market,
the properties of the function $x\rightarrow {\mathcal{A}}(x)$ are solely
related to distributional properties of final wealth and do not depend on
particular market conditions. This observation is implicit in the following
theorem.

\begin{theo}[DARA in a Black-Scholes market]
\label{SUFF-2} Consider an investor who targets some cdf $F$ for his
terminal wealth. In a Black-Scholes market the investor has decreasing
absolute risk aversion if and only if 
\begin{equation}
\frac{f(F^{-1}(p))}{\phi (\Phi ^{-1}(p))}\text{ is strictly decreasing on }%
(0,1)  \label{daraBS}
\end{equation}%
or, equivalently, 
\begin{equation}
F^{-1}(\Phi (x))\text{ is strictly convex on }\mathbb{R},  \label{daraBSalt}
\end{equation}%
where $\phi (\cdot )$ and $\Phi (\cdot )$ are the density and the cdf of a
standard normal distribution.
\end{theo}

The proof of this Theorem is given in Appendix \ref{SUFF-2app}. When $F$ is
the distribution of a normal random variable, the ratio (\ref{daraBS})
becomes constant. This confirms the well-known fact that the demand for a
normally distributed final wealth in a Black-Scholes market is tied to a
constant absolute risk aversion (see Section \ref{NEXP} for a formal proof).
In a Black-Scholes setting, the property of DARA remains invariant to
changes in the financial market. This is, however, not true in a general
market, where risk aversion, demand for a particular distribution and
properties of the financial market are intertwined.

If the final wealth $W$ has a lognormal distribution $F$, then one has $%
F^{-1}(\Phi (x))=\exp (x)$. Since the exponential function is clearly
convex, Theorem \ref{SUFF-2} implies that, in a Black-Scholes market, the
demand for a lognormal distribution corresponds to DARA preferences. One can
also readily show that the exponential distribution also corresponds to DARA
preferences in this setting. Effectively, from \eqref{daraBSalt}, we only
need to show that the survival function $1-\Phi (x)$ of a standard normal
random variable is log-concave.\footnote{\label{foof}This is well-known in
the literature and can be seen as a direct consequence of a more general
result attributed to Pr{\'e}kopa \citeyear{Pr}, who shows that the
differentiability and log-concavity of the density implies log-concavity of
the corresponding distribution and survival function. It is clear that a
normal density is log-concave and thus also its survival function.} 

%

Consider for all $x\in \mathbb{R}$ such that $F(x)<1$, the hazard function $%
h(x):=\frac{f(x)}{1-F(x)}$, which is a useful device for studying
heavy-tailed properties of distributions. The hazard function for an
exponentially distributed random variable rate function is clearly constant,
so that a non-increasing hazard function reflects a distribution that is
heavier-tailed than an exponentially distributed random variable. It is
therefore intuitive that investors exhibiting a demand for distributions
with a non-increasing hazard function exhibit DARA preferences. The
following theorem makes this precise.

%
%
%
%
%
%

\begin{theo}[Sufficient conditions for DARA]
\label{T2} Consider an investor who targets some cdf $F$ for her terminal
wealth. Denote its density by $f$. If the hazard function $h(x)$ is
non-increasing (resp. non-increasing for $x>k$) or, equivalently, if $1-F(x)$
is log-convex (resp. log-convex for $x>k$), then the investor has decreasing
absolute risk aversion (resp. asymptotic DARA).
\end{theo}



We remark from the proof (Appendix \ref{T2app}) that a random variable with
non-increasing hazard function $h(x)$ must assume values that are almost
surely in an interval $[a,\infty )$ where $a\in \mathbb{R}$. A lognormally
distributed random variable thus has no non-increasing hazard function (but
still satisfies the DARA property). 

\section{Distributions and Corresponding Utility Functions\label{S4}}

We end this paper with a few examples illustrating the correspondence
between the distribution of final wealth and the utility function. 

\subsection{Normal Distribution and Exponential Utility\label{NEXP}}

Let $F$ be the distribution of a normal random variable with mean $M$ and
variance $\Sigma^{2}$. Then, from Theorem \ref{th1}, the utility function
explaining this distribution writes as 
\begin{equation}
u(x)=-\frac{\Sigma a}{\theta \sqrt{T}}\exp \left( -\frac{x\theta }{\Sigma }%
\sqrt{T}\right)  \label{Normalutility}
\end{equation}%
where $a=\exp \left( \frac{M\theta \sqrt{T}}{\sigma }-rT-\frac{\theta ^{2}T}{%
2}\right) $ and $\theta =\frac{\mu -r}{\sigma }$. This is essentially the
form of an exponential utility function with constant absolute risk aversion 
\begin{equation*}
{\mathcal{A}}(x)=\frac{\theta \sqrt{T}}{\Sigma }.
\end{equation*}%
Note that the absolute risk aversion is constant and inversely proportional
to the volatility $\Sigma $ of the distribution. A higher volatility of the
optimal distribution of final wealth corresponds to a lower absolute risk
aversion. This is consistent with Dybvig and Wang \citeyear{DW}, who show
that lower risk aversion leads to a larger payoff in the sense of SSD.%
\footnote{%
In case of two normal distributions with equal mean, increasing SSD is
equivalent to increasing variance.} 
Reciprocally, consider the following exponential utility function defined
over $\mathbb{R}$, 
\begin{equation}
U(x)=-\exp(-\gamma x),  \label{EXPutility}
\end{equation}%
where $\gamma $ is the risk aversion parameter and $x\in\mathbb{R}$. This
utility has constant absolute risk aversion ${\mathcal{A}}(x)=\gamma $. The
optimal wealth obtained with an initial budget $X_0$ is given by 
\begin{equation}
X_{T}^{\star }=X_0e^{rT}-\frac{\theta }{\gamma \sigma }\left( r-\frac{\sigma
^{2}}{2}\right) T+\frac{\theta }{\gamma \sigma }\ln \left( \frac{S_{T}}{S_{0}%
}\right),  \label{EXPpayoff}
\end{equation}%
where $\theta =\frac{\mu -r}{\sigma }$ is the instantaneous Sharpe ratio for
the risky asset $S$, and thus $X_{T}^{\star }$ follows a normal distribution 
$\mathcal{N}\left( X_0e^{rT}+\frac{\theta }{\gamma \sigma }(\mu -r)T,\left( 
\frac{\theta }{\gamma }\right) ^{2}T\right) $. 

\subsection{Lognormal Distribution and CRRA and HARA Utilities}

The HARA utility is a generalization of the CRRA utility and is given by 
\begin{equation}
U(x)=\frac{1-\gamma }{\gamma }\left( \frac{ax}{1-\gamma }+b\right) ^{\gamma
},  \label{HARAutility}
\end{equation}%
where $a>0$, $b+\frac{ax}{1-\gamma }>0$. For a given parametrization, this
restriction puts a lower bound on $x$ if $\gamma <1$ and an upper bound on $%
x $ when $\gamma >1$. If $-\infty <\gamma <1$, then this utility displays
DARA. In the case $\gamma \rightarrow 1$, the limiting case corresponds to a
linear utility. In the case $\gamma \rightarrow 0$, the utility function
becomes logarithmic: $U(x)=\log (x+b)$. Its absolute risk aversion is 
\begin{equation}
{\mathcal{A}}(x)=a\left( \frac{ax}{1-\gamma }+b\right) ^{-1}.
\label{HARAabsriskaversion}
\end{equation}%
The optimal wealth obtained with an initial budget $X_0$ is given by 
\begin{equation*}
X_{T}^{\star }=C\left( \frac{S_{T}}{S_{0}}\right) ^{\frac{\theta }{\sigma
(1-\gamma )}}-\frac{b(1-\gamma )}{a}
\end{equation*}%
where $C=\frac{X_0e^{rT}+\frac{b(1-\gamma )}{a}}{\exp \left( \frac{\theta }{%
\sigma (1-\gamma )}\left( r-\frac{\sigma ^{2}}{2}\right) T+\left( \frac{%
\theta }{1-\gamma }\right) ^{2}\frac{T}{2}\right) }$. Its cdf is 
\begin{equation*}
F_{HARA}(y)=\Phi \left( \frac{\ln \left( \frac{y+\frac{b(1-\gamma )}{a}}{C}%
\right) +\frac{\theta }{\sigma (\gamma -1)}\left( \mu -\frac{\sigma ^{2}}{2}%
\right) T}{\frac{\theta }{\gamma -1}\sqrt{T}}\right) .
\end{equation*}%
Observe that the optimal wealth for a HARA utility is a lognormal
distribution translated by a constant term.

When $b=0$, the HARA utility reduces to CRRA and the optimal wealth is
LogNormal $\mathcal{LN} (M,\Sigma ^{2})$ over $\mathbb{R}^{+}$. We showed in
Section \ref{S2} that the utility function explaining this distribution is a
CRRA utility function. It has decreasing absolute risk aversion 
\begin{equation*}
{\mathcal{A}}(x)=\frac{\theta \sqrt{T}}{x\Sigma }\ \hbox{where}\ \theta=%
\frac{\mu-r}{\sigma}.
\end{equation*}

\subsection{Exponential Distribution}

Consider an exponential distribution with cdf $F(x)=1-e^{-\lambda x}$ where $%
\lambda >0$. The utility function in \eqref{Utility1} in Theorem \ref{th1}
cannot be obtained in closed-form and does not correspond to a well-known
utility. 
The coefficient of absolute risk aversion is given by 
\begin{equation*}
{\mathcal{A}}(x)=\theta \lambda \sqrt{2\pi T}\exp \left( -\lambda x+\frac{1}{%
2}\left[ \Phi ^{-1}\left( e^{-\lambda x}\right) \right] ^{2}\right) .
\end{equation*}%
%
%
%
%
%
%
%
%
%
%
%
%
%
%
%
%
%
%
%
%
%
%
%
%
%
%
%
%
%
%
%
%
%
%
%
%
%
%
%
%
%
%
%
%
%
%
%
%
%
%
%
%
%
%
%
%
%
We have already shown that ${\mathcal{A}}(x)$ is decreasing and thus that
the exponential distribution corresponds to a utility function exhibiting
DARA.

\subsection{Pareto Distribution}

Consider a Pareto distribution with scale $m>0$ and shape $\alpha >0$,
defined on $[m,+\infty )$. Its pdf is $f(x)=\alpha \frac{m^{\alpha }}{%
x^{\alpha +1}}\mathds{1}_{x\geq m}$. 
The coefficient of absolute risk aversion for $x\geq m$ is given by 
\begin{equation*}
{\mathcal{A}}(x)=\frac{\theta \alpha m^{\alpha }\sqrt{2\pi T}}{x^{\alpha +1}}%
\exp \left( \frac{1}{2}\left[ \Phi ^{-1}\left( \frac{m^\alpha}{x^\alpha}%
\right) \right] ^{2}\right).
\end{equation*}%
%
%
%
%
%
%
%
%
%
%
%
%
%
%
%
%
%
%
%
%
%
%
%
%
%
%
%
%
%
%
%
%
%
%
%
%
%
%
%
%
%
%
%
%
%
%
%
%
%
%
%
%
%
%
%
%
%
Bergstrom and Bagnoli \citeyear{BeBa} show that $1-F(x)$ is log-convex.
Theorem \ref{T2} thus implies that the Pareto distribution corresponds to
DARA preferences.

\section*{Conclusions}

Assuming investors' preferences satisfy FSD, our results can be used to
non-parametrically estimate the utility function and the risk aversion of an
investor, based solely on knowledge of the distribution of her final wealth
(when optimally investing) and of the distribution of the pricing kernel
(e.g., estimated in a non-parametric way, as per A{ï}t-Sahalia and Lo %
\citeyear{AL}). Another application consists in inferring the utility
function that rationalizes optimal investment choice in non-expected utility
frameworks (e.g., cumulative prospect theory or rank dependent utility
theory).


Our results suggest inherent deficiencies in portfolio selection within a
decision framework that satisfies FSD (equivalently with a law-invariant
non-decreasing objective function). When the market exhibits a positive risk
premium, Bernard, Boyle and Vanduffel \citeyear{BBV} show that for every put
contract there exists a derivative contract that yields the same
distribution at a strictly lower cost but without protection against a
market crash. Then, the demand for standard put contracts can only be
rationalized in a decision framework that violates FSD and in which
background risk as a source of state-dependent preferences comes into play.
Similarly, insurance contracts provide protection against certain losses and
provide a payout when it is needed. For this reason, they often appear more
valuable to customers than (cheaper) financial payoffs with the same
distribution (Bernard and Vanduffel \citeyear{BV}). Thus, it would be useful
to seek to develop decision frameworks in which FSD can be violated, for
example considering ambiguity on the pricing kernel, Almost Stochastic
Dominance (ASD) (Levy \citeyear{Lbook}, chapter 13) or state-dependent
preferences (Bernard, Boyle and Vanduffel \citeyear{BBV}, Chabi-Yo, Garcia
and Renault \citeyear{chabi2008state}). \newpage


\appendix

\section{Appendix: Proofs of Theorems}

\subsection{Proof of Proposition \protect\ref{PR0} on page \protect\pageref%
{PR0}\label{A0}}

\proof Using the concavity of $U(\cdot )$, it is clear that for $i\in
\{1,2,...N\}$, 
\begin{equation}  \label{keykey}
\forall x>x_{i}^{\ast },\ \frac{U(x)-U(x_{i}^{\ast })}{x-x_{i}^{\ast }}\leq
U^{\prime }(x_{i}^{\ast }),\quad \quad \forall x<x_{i}^{\ast },\ \frac{%
U(x)-U(x_{i}^{\ast })}{x-x_{i}^{\ast }}\geq U^{\prime }(x_{i}^{\ast }).
\end{equation}
Thus for all $x\in \mathbb{R}$, and $i\in \{1,2,...,N\}$, using \eqref{key}
and \eqref{keykey}, 
\begin{equation*}
U(x_{i}^{\ast })-\xi _{i}x_{i}^{\ast }=U(x_{i}^{\ast })-U^{\prime
}(x_{i}^{\ast })x_{i}^{\ast }\geq U(x)-U^{\prime }(x_{i}^{\ast })x=U(x)-\xi
_{i}x.
\end{equation*}%
Therefore, for all $\omega \in \Omega $, and terminal consumption $X$, 
\begin{equation}
U(X^{\ast }(\omega ))-\xi (\omega )X^{\ast }(\omega )\geq U(X(\omega ))-\xi
(\omega )X(\omega ).  \label{of5}
\end{equation}%
Since $X^{\ast }$ solves \eqref{RE}, $E[\xi X^{\ast }]=X_{0}$. Then, for all
terminal consumption $X$ such that $E[\xi X]=X_{0}$ (that is, all attainable
consumption with the given budget $X_{0}$), after taking the expectation of %
\eqref{of5}, $E[U(X^{\ast })]\geq E[U(X)]$. In other words, $X^{\ast }$ also
solves \eqref{RE2}.\hfill $\Box $

\subsection{Proof of Theorem \protect\ref{LI} on page \protect\pageref{LI} 
\label{appLI}.}

Assume that $V(\cdot )$ is non-decreasing and law-invariant and consider $X$
and $Y$ with respective distributions $F_{X}$ and $F_{Y}$. Assume that $%
X\prec _{fsd}Y$. One has, for all $x\in(0,1)$, $F_{X}^{-1}(x)\leq
F_{Y}^{-1}(x)$. Let $U$ be a uniform random variable over $(0,1)$. Hence, $%
F_{X}^{-1}(U)\leq F_{Y}^{-1}(U)$ a.\thinspace s. . Moreover, $X\sim
F_{X}^{-1}(U)$ and $Y\sim F_{Y}^{-1}(U)$ so that $V(X)=V(F_{X}^{-1}(U))\leq
V(F_{Y}^{-1}(U))=V(Y).$ Thus $V(\cdot )$ satisfies FSD. Conversely, assume
that $V(\cdot )$ satisfies FSD. If $X\sim Y$, then $X\prec _{fsd}Y$ and $%
Y\prec _{fsd}X$, which implies $V(X)=V(Y)$ and thus law invariance. Clearly,
if $X\leq Y$ a.\thinspace s. then $X\prec _{fsd}Y$ and $V(X)\leq V(Y).$%
\hfill $\Box $

\subsection{Proof of Lemma \protect\ref{lemma1} on page \protect\pageref%
{lemma1} \label{Lem1proof}}

Given $\omega \in \Omega $, consider the following auxiliary problem 
\begin{equation}
\max_{y\in (a,b)}\ \left\{ U(y)-\lambda \xi _{T}(\omega )y\right\}
\label{axi}
\end{equation}%
with $\lambda \in \mathbb{R}^{+}$. This is an optimization over the interval 
$(a,b)$ of a concave function. The first-order conditions imply that the
optimum $y^{\ast }$ is at $U^{\prime }(y)-\lambda \xi _{T}(\omega )=0,$
i.e., $y^{\ast }=[U^{\prime }]^{-1}(\lambda \xi _{T}(\omega ))$. For each $%
\omega \in \Omega $, define the random variable $Y_{\lambda }^{\star }$ by $%
Y_{\lambda }^{\star }(\omega )=y^{\ast }$ so that 
\begin{equation*}
Y_{\lambda }^{\star }(\omega )=[U^{\prime }]^{-1}(\lambda \xi _{T}(\omega )).
\end{equation*}%
Choose $\lambda ^{\ast }>0$ such that $E[\xi _{T}Y_{\lambda ^{\ast }}^{\star
}]=X_{0}$. The existence of $\lambda ^{\ast }$ is ensured by the conditions
imposed on $U(\cdot )$, by continuity of $\lambda \mapsto E[\xi
_{T}Y_{\lambda }^{\star }]$ and the fact that the budget $X_{0}\in (E[\xi
_{T}a],E[\xi _{T}b])$. For every final wealth $X_{T}$ that satisfies the
budget constraint of \eqref{EUpb}, we have by construction, 
\begin{equation*}
U(X_{T}(\omega ))-\lambda ^{\ast }\xi _{T}(\omega )X_{T}(\omega )\leq
U(Y_{\lambda }^{\star }(\omega ))-\lambda ^{\ast }\xi _{T}(\omega
)Y_{\lambda }^{\star }(\omega )
\end{equation*}%
since $Y_{\lambda }^{\star }(\omega )$ is the optimal solution to \eqref{axi}%
. Now, take the expectation on both sides of the above inequality. As $E[\xi
_{T}X_{T}]=E[\xi _{T}Y_{\lambda ^{\star }}]=X_{0}$, we obtain 
\begin{equation*}
E[U(X_{T})]\leq E[U(Y_{\lambda ^{\ast }}^{\star })],
\end{equation*}%
which ends the proof that $Y_{\lambda ^{\ast }}^{\star }$ is optimal for
Problem \eqref{EUpb}. Since $[U^{\prime }]^{-1}(\cdot )$ is strictly
decreasing and $\xi _{T}$ has a continuous distribution, it is clear that $%
Y_{\lambda ^{\ast }}^{\star }$ has a continuous and strictly increasing
distribution on $(a,b)$ with $F(a^{+})=0$ and $F(b^{-})=1$. \hfill $\Box $

\subsection{Proof of Theorem \protect\ref{thD} on page \protect\pageref{thD}}

$(vi)$ and $(vii)$ are equivalent because of Theorem \ref{LI}. Given that $%
X_{T}$ has distribution $F$, we have that $(ii)$, $(iii)$, $(iv)$ and $(v)$
are equivalent as a consequence of Lemma \ref{th0} and of the earlier work
of Dybvig \citeyear{dybvigJoB,dybvigRFS} and Bernard et al. \citeyear{BBV}
on cost-efficiency. In addition, it has already been noted that a solution
to \eqref{EUopt} must be cost-efficient (Lemma \ref{th0}). Thus $(vi)$ or $%
(vii)$ implies (ii) 
From Theorem \ref{th1}, it is clear that a cost-efficient payoff is the
solution to an expected utility maximization (so $(ii)$ implies $(i)$).
Finally observe that if $X_{T}$ is rationalizable by the standard expected
utility maximization then $(vii)$ also holds true (so $(i)$ implies $(vii)$%
). We have proved that the seven first statements are equivalent. Let's
prove the equivalence with $(viii)$. Assume $(viii)$ and observe that if $%
V(\cdot )$ satisfies SSD\footnote{$X$ is smaller then $Y$ for SSD, $X\prec
_{ssd}Y$, when for all $u$ concave and non-decreasing, $E[u(X)]\leq E[u(Y)]$.%
} then $V(\cdot )$ satisfies also FSD\footnote{$X$ is smaller then $Y$ for
FSD, $X\prec _{fsd}Y$, when for all $u$ non-decreasing, $E[u(X)]\leq E[u(Y)]$%
.}, thus $(viii)$ implies $(vi).$ We have also proved that $(vi)$ is
equivalent to $(i)$. But then $(i)$ implies $(viii)$ because $X\rightarrow
V(X):=E[u(X)]$ where $u$ is concave in $\mathcal{U}_{(a,b)}$ satisfies
SSD.\hfill $\Box $

\subsection{Proof of Lemma \protect\ref{lemma1bis} on page \protect\pageref%
{lemma1bis} \label{lemma1bisProof}}

Take $\widetilde{U}\in\widetilde{\mathcal{U}}_{(a,b)}$ (Definition \ref{DEFI}%
). 
Note that $U(x)$ is concave and strictly increasing and $(a,b)\subset 
\mathbb{R}$, which implies that $\widetilde{U}(x)$ is continuous and
strictly increasing on $(a,b),$ Hence, the 
left and right derivatives of $\widetilde{U}$ exist at each point of $(a,b)$
and are equal to each other, except in a countable number of points where $%
\widetilde{U}$ is not differentiable; see also Proposition 17, Chapter 5,
Section 5 of Royden \citeyear{Royden}). We can then conclude that $%
\widetilde{U}^{\prime }$ is left-continuous and decreasing on $(a,\infty )$
with discontinuities in a set $(x_{i})_{i\in I}$ with $x_{i}\in (a,b]$ and $%
I\subset 
\mathbb{N}
$ (note that $b$ can also be a point of non-differentiability)$.$ There is
also a countable number of non-empty open intervals of $\mathbb{R}$ on which 
${\tilde{U}}^{\prime }$ is constant (they are disjoint open intervals of $%
\mathbb{R}$ and thus countable). Define $\{y_{i}\}_{i\in J}$ the countable
set of values taken by $\widetilde{U}^{\prime }$ on these intervals (with $%
J\subset 
\mathbb{N}
)$. Let $\lambda >0$ and define 
\begin{equation*}
A=\bigcup_{i\in J}\{\omega \in \Omega \text{ }|\text{ }\lambda \xi
_{T}(\omega )=y_{i}\}.
\end{equation*}%
$P(A)=0$ as it is a countable union of sets, each with zero probability
(given that $\xi _{T}$ is continuously distributed).

The proof consists further of three steps:

\noindent \underline{Step 1:} Given $\lambda >0$ and $\omega \in \Omega
\backslash A$, we show below that the a.s. unique optimum of 
\begin{equation}
\max_{y\in \lbrack a,b]}\ \left\{ \tilde{U}(y)-\lambda \xi _{T}(\omega
)y\right\}
\end{equation}%
is equal to $Y_{\lambda }^{\star }(\omega ):=\left[ {\tilde{U}}^{\prime }%
\right] ^{-1}(\lambda \xi _{T}(\omega ))\in 
\mathbb{R}
,$ where the inverse of ${\tilde{U}}^{\prime }$ is defined by \eqref{TU},

\noindent \underline{Step 2:} The a.s. optimum of 
\begin{equation}
\max_{y\in \mathbb{R}}\ \left\{ \tilde{U}(y)-\lambda \xi _{T}(\omega
)y\right\}  \label{aux22}
\end{equation}
is then also equal to $Y_{\lambda }^{\star }(\omega )$ too because 
\begin{equation*}
\forall z>b,\quad \tilde{U}(z)-\lambda \xi _{T}(\omega )z<\tilde{U}%
(b)-\lambda \xi _{T}(\omega )b
\end{equation*}%
and%
\begin{equation*}
\forall z<a,\quad \tilde{U}(z)-\lambda \xi _{T}(\omega )z<\tilde{U}%
(a)-\lambda \xi _{T}(\omega )a,
\end{equation*}%
since for $z>b$, $\tilde{U}(b)=\tilde{U}(z),$ and $\forall z<a,$ $\tilde{U}%
(z)-\lambda \xi _{T}(\omega )z=$ $-\infty$ and $\lambda >0$.

\noindent \underline{Step 3:} If there exists $\lambda >0$ such that the
pathwise optimum $Y_{\lambda }^{\star }$ satisfies the budget constraint $%
E[\xi _{T}Y_{\lambda }^{\star }]=X_{0}$, then $Y_{\lambda }^{\star }$ solves
Problem (\ref{eu2b}) (note that $P(A)=0)$. The existence of $\lambda $ is
guaranteed thanks to the budget constraint $X_{0}\in (E[\xi _{T}a],E[\xi
_{T}b])$, the property that the cost as a function of $\lambda $ is
continuous on $(0,\infty )$, and the facts that if $\lambda \rightarrow
+\infty $ the limit of the pseudo inverse of the left derivative is $a$ and
that if $\lambda \rightarrow 0$, the limit of the pseudo inverse of the left
derivative is $b$.

To complete the proof of Lemma \ref{lemma1bis}, the only remaining step to
prove is Step 1. For $\lambda >0$, we have that $\lambda \xi _{T}(\omega )>0$%
. 
Denote by 
\begin{equation*}
x^{\ast }:=\left[ {\tilde{U}}^{\prime }\right] ^{-1}(\lambda \xi _{T}(\omega
))
\end{equation*}%
where $\left[ {\tilde{U}}^{\prime }\right] ^{-1}$ is defined in \eqref{TU}.
Specifically it is $\left[ {\tilde{U}}^{\prime }\right] ^{-1}(y)=\inf
\left\{ x\in (a,b)\ |\ {\tilde{U}}^{\prime }(x)\leq y\right\} $, so that it
is clear that $x^{\ast }\in \lbrack a,b]$. Note that $x^{\ast }\in 
\mathbb{R}
.$ Indeed, if $a=-\infty $ then $x^{\ast }>a$ because ${\tilde{U}}^{\prime
}(a)=\infty .$ Similarly if $b=+\infty $, then $x^{\ast }<b$ because ${%
\tilde{U}}^{\prime }(b)=0.$ Thus, $x^{\ast }\in \mathbb{R}$ in all cases.
Let us prove that $x^{\ast }$ is an optimum. As $\widetilde{U}^{\prime }$ is
left-continuous, $\widetilde{U}^{\prime }(x^{\ast })\geq \lambda \xi
_{T}(\omega )$ and we consider the following two cases:\newline
\underline{Case 1:} $\widetilde{U}^{\prime }(x^{\ast })=\lambda \xi
_{T}(\omega ).$ Then for $x\in (a,x^{\ast })$, 
\begin{equation*}
\begin{array}{rl}
\tilde{U}(x^{\ast })-\lambda \xi _{T}(\omega )x^{\ast }-(\tilde{U}%
(x)-\lambda \xi _{T}(\omega )x) & =\left[ \frac{\tilde{U}(x^{\ast })-\tilde{U%
}(x)}{x^{\ast }-x}-\lambda \xi _{T}(\omega )\right] (x^{\ast }-x) \\ 
& \geq (\lambda \xi _{T}(\omega )-\lambda \xi _{T}(\omega ))(x^{\ast }-x)=0,%
\end{array}%
\end{equation*}%
where we use the concavity of $\tilde{U}(x)$ (specifically, the fact that
the slope $\frac{\tilde{U}(x^{\ast })-\tilde{U}(x)}{x^{\ast }-x}$ is larger
than $\tilde{U}^{\prime }(x^{\ast })=\lambda \xi _{T}(\omega )$). For $x\in
(x^{\ast },b)$, using the concavity of $\tilde{U}(x)$ again, we have that 
\begin{equation*}
\begin{array}{rl}
\tilde{U}(x^{\ast })-\lambda \xi _{T}(\omega )x^{\ast }-(\tilde{U}%
(x)-\lambda \xi _{T}(\omega )x) & =\left[ -\frac{\tilde{U}(x)-\tilde{U}%
(x^{\ast })}{x-x^{\ast }}+\lambda \xi _{T}(\omega )\right] (x-x^{\ast }) \\ 
& \geq (-\lambda \xi _{T}(\omega )+\lambda \xi _{T}(\omega ))(x-x^{\ast })=0.%
\end{array}%
\end{equation*}%
For all $x\in (a,b)$ we have proved that $\tilde{U}(x^{\ast })-\lambda \xi
_{T}(\omega )x^{\ast }-(\tilde{U}(x)-\lambda \xi _{T}(\omega )x)\geq 0$. The
same inequality also holds for $x=a$ and $x=b$ (we take take limits if $%
a=-\infty $ or $b=\infty $)%

\underline{Case 2:} $\widetilde{U}^{\prime }(x^{\ast })>\lambda \xi
_{T}(\omega ).$ This implies that $x^{\ast }$ is at some point of
non-differentiability $x_{i},$ $x^{\ast }=x_{i}$, and the left derivative of 
$\widetilde{U}$ in $x_{i}$ is strictly larger than its right derivative,
which we denote by $a_{i}^{(r)}$. $\widetilde{U}^{\prime }(x_{i})=\widetilde{%
U}^{\prime }(x^{\ast })>\lambda \xi _{T}(\omega )>a_{i}^{(r)}.$ Observe that
when $a<x<x_{i}$, by concavity of $\tilde{U}(x)$, 
\begin{equation*}
\begin{array}{rl}
\tilde{U}(x_{i})-\lambda \xi _{T}(\omega )x_{i}-\tilde{U}(x)+\lambda \xi
_{T}(\omega )x & =\left( \frac{\tilde{U}(x_{i})-\tilde{U}(x)}{x_{i}-x}%
-\lambda \xi _{T}(\omega )\right) (x_{i}-x) \\ 
& \geq (\widetilde{U}^{\prime }(x_{i})-\lambda \xi _{T}(\omega
))(x_{i}-x)\geq 0.%
\end{array}%
\end{equation*}%
When $b>x>x_{i}$, using the concavity of $\tilde{U}(x)$ again, 
\begin{equation*}
\begin{array}{rl}
\tilde{U}(x_{i})-\lambda \xi _{T}(\omega )x_{i}-\tilde{U}(x)+\lambda \xi
_{T}(\omega )x & =\left( -\frac{\tilde{U}(x)-\tilde{U}(x_{i})}{x-x_{i}}%
+\lambda \xi _{T}(\omega )\right) (x-x_{i}) \\ 
& \geq (-a_{i}^{(r)}+\lambda \xi _{T}(\omega ))(x-x_{i})\geq 0, \\ 
& \geq (-\widetilde{U}^{\prime }(x_{i})+\lambda \xi _{T}(\omega
))(x-x_{i})\geq 0.%
\end{array}%
\end{equation*}%
For all $x\in (a,b),$ and thus for all $x\in \lbrack a,b]$, we have proved
that $\tilde{U}(x_{i})-\lambda \xi _{T}(\omega )x_{i}-(\tilde{U}(x)-\lambda
\xi _{T}(\omega )x)\geq 0$.

We can thus conclude that $x^{\ast }$ is an optimum. It is also unique
unless $\tilde{U}(y)-\lambda \xi _{T}(\omega )$ is flat at $x^{\ast }$, but
by assumption $\omega \in \Omega \backslash A$ and thus the slope of $%
y\mapsto \tilde{U}(y)-\lambda \xi _{T}(\omega )y$ can never be $0$ on an
interval that is not reduced to a point. This ends the proof of Step 1 and
thus Lemma \ref{lemma1bis} is proved. \hfill \mbox{\quad$\square$}

\subsection{Proof of Theorem \protect\ref{th3} on page \protect\pageref{th3}.%
}

\noindent\underline{First step: Verify that $\widetilde U$ in %
\eqref{FormulaUtility2} belongs to $\widetilde{\mathcal{U}}_{(a,b)}$ for
some real numbers $a$ and $b$.}

Define $a=\inf \{x\ |\ F(x)>0\}$ and $b=\sup \{x\ |\ F(x)<1\}$. Let us prove
that $\widetilde{U}\in \widetilde{\mathcal{U}}_{(a,b)}.$

Observe that $F_{\xi _{T}}(1-F(y))>0$ for all $y\in (a,b)$ (because $\xi
_{T} $ is continuously distributed) and thus $\widetilde{U}$ is strictly
increasing on $(a,b)$. Its left derivative for $y\in (a,b)$ is given by 
\begin{equation}
\widetilde{{U}}^{\prime }(y)=F_{\xi _{T}}^{-1}\left( 1-F(y)\right) .
\label{DER}
\end{equation}%
It is clear that it is non-increasing on $(a,b)$ and thus $\widetilde{U}$ is
concave on $(a,b).$ $\widetilde{U}(a)$ and $\widetilde{U}(b)$ are
well-defined (potentially in $\bar{\mathbb{R}}$).

Since $F_{\xi _{T}}^{-1}(1)=+\infty $, $F_{\xi _{T}}^{-1}(0)=0$ and using
the convention that the integral of $+\infty $ between $x_{1}$ and $x_{2}$
is $+\infty $ when $x_{1}<x_{2}$ and $-\infty $ when $x_{1}>x_{2}$ then $%
\widetilde{U}(y)=-\infty $ for $y<a$ and $\widetilde{U}(y)=\widetilde{U}(b)$
for $y>b$. We have that 
\begin{equation*}
\widetilde{U}^{\prime }(a)=\lim_{y\searrow a}F_{\xi _{T}}^{-1}\left(
1-F(y)\right)
\end{equation*}%
When $a=-\infty $, then $\widetilde{U}^{\prime }(y)\rightarrow +\infty $
when $y\rightarrow a$ because $F_{\xi _{T}}^{-1}(1)=+\infty .$ When $%
b=\infty $, then $\widetilde{U}^{\prime }(y)\rightarrow 0$ when $%
y\rightarrow b$ because $F_{\xi _{T}}^{-1}(0)=0.$

\noindent\underline{Second step: Apply Lemma \ref{lemma1bis} and compute the
distribution of $X_T^*$.} Using Lemma \ref{lemma1bis}, the optimal solution
to Problem \eqref{eu2b} is 
\begin{equation*}
X_{T}^{\star }:=\left[ \widetilde{U}^{\prime }\right] ^{-1}\left( \lambda
^{\ast }\xi _{T}\right)
\end{equation*}
where $\lambda ^{\ast }$ $>0$ is such that $E\left[ \xi _{T}X_{T}^{\star } %
\right] =X_0$, and for $y>0$, the pseudo inverse of $\widetilde U^\prime$ is
defined by \eqref{TU}. 
Given its expression in \eqref{DER}, $\left[ {\tilde{U}}^{\prime }\right]
^{-1}(y)=\inf \left\{ x\in (a,b)\ |\ F_{\xi_T}^{-1}\left(1-F(x)\right)\leq
y\right\}$, which can be rewritten as 
\begin{equation*}
\left[ {\tilde{U}}^{\prime }\right] ^{-1}(y)=\inf \left\{ x\in (a,b)\ |\
F(x)\geq 1-F_{\xi_T}(y)\right\}=F^{-1}(1-F_{\xi_T}(y)).
\end{equation*}
where $0<1-F_{\xi_T}(y)<1$ because $\xi_T$ has a positive density over
positive real numbers and $y>0$. Thus 
\begin{equation*}
X_{T}^{\star }=F^{-1}(1-F_{\xi_T}(\lambda^*\xi_T))
\end{equation*}
For $\lambda^*=1$, $X_{T}^{\star }$ has the distribution $F$ and the right
cost $X_0$. We have thus proved the existence and the form of the
generalized utility function that explains the demand for $F$.

\noindent \underline{Third step: Uniqueness of $\widetilde{U}$.} Assume that
there exists $a_{2}$ and $b_{2}$ and another generalized utility $\widetilde{%
U}_{2}\in \widetilde{\mathcal{U}}_{(a_{2},b_{2})}$ such that the optimum has
distribution $F$. From Lemma \ref{th0}, there is a unique cost-efficient
payoff with cdf $F$ (i.e., it is given as $F^{-1}(1-F_{\xi _{T}}(\xi _{T}))$%
). From Lemma \ref{lemma1bis}, there is a unique optimum for a given
generalized utility $\widetilde{U}_{2}$ for the generalized expected utility
maximization given by $\left[ \widetilde{U}_{2}^{\prime }\right] ^{-1}\left(
\lambda ^{\ast }\xi _{T}\right) $, so we get that 
\begin{equation*}
F^{-1}(1-F_{\xi _{T}}(\xi _{T}))=\left[ \widetilde{U}_{2}^{\prime }\right]
^{-1}\left( \xi _{T}\right) \ \ a.s.
\end{equation*}%
Thus, 
\begin{equation*}
\left[ \widetilde{U}^{\prime }\right] ^{-1}\left( \xi _{T}\right) =\left[ 
\widetilde{U}_{2}^{\prime }\right] ^{-1}\left( \xi _{T}\right) \ \ a.s.
\end{equation*}%
We must have $a=a_{2},$ otherwise, one would be finite while the other side
is infinite. We must also have $b=b_{2},$ otherwise, one would be 0 while
the other one is positive. Finally, two continuous functions with the same
left derivatives at any point on $(a,b)$ must be equal up to a linear
transformation. The uniqueness of the generalized utility is thus proved.
\hfill $\Box $

%

\subsection{Proof of Theorem \protect\ref{RAVERSION} on page \protect\pageref%
{RAVERSION}\label{appRAVERSION}}

Observe that $F_{\xi_T}^{-1}(y)=\exp(-G^{-1}(1-y))$ so that using the
expression for $U$ obtained in \eqref{Utility1}, 
\begin{equation*}
U^{\prime }(x)=F_{\xi _{T}}^{-1}(1-F(x))=\exp (-G^{-1}(F(x)))
\end{equation*}%
and 
\begin{equation*}
U^{\prime \prime }(x)=U^{\prime }(x)\frac{-f(x)}{g(G^{-1}(F(x)))}.
\end{equation*}%
The stated expression for the absolute risk aversion coefficient ${\mathcal{A%
}}(x)$ is then obtained, as ${\mathcal{A}}(x)=-\frac{U^{\prime \prime }(x)}{%
U^{\prime }(x)}$. It is well-defined when $F(x)\in(0,1)$ because $g(x)>0$
for all $x\in\mathbb{R}$ and $G^{-1}$ exists on $(0,1)$. The expression for
the relative risk aversion coefficient ${\mathcal{R}}(x)$ follows
immediately. \hfill $\Box $

\subsection{Proof of Theorem \protect\ref{SUFF} on page \protect\pageref%
{SUFF} \label{SUFFapp}}

Denote by $\mathbb{F}$ the domain where ${\mathcal{A}}(x)$ is well-defined.
To assess the DARA property, we study when $\ln ({\mathcal{A}}(x))$ is
strictly decreasing for $x\in \mathbb{F}$. For all $x\in \mathbb{F}$, it is
clear that $x\mapsto \ln ({\mathcal{A}}(x))$ is decreasing if and only if $%
p\mapsto \ln ({\mathcal{A}}(F^{-1}(p)))$ is decreasing on $(0,1)$. Using
expression \eqref{Absriskaversion} for ${\mathcal{A}}(x)$ for $x\in \mathbb{F%
}$, and differentiating with respect to $p,$ one has that the derivative is
negative if and only if for $p\in (0,1)$, 
\begin{equation}
\frac{g^{\prime }(G^{-1}(p))}{g^{2}(G^{-1}(p))}>\frac{f^{\prime }(F^{-1}(p))%
}{f^{2}(F^{-1}(p))}.  \label{DARACondition}
\end{equation}%
Consider next the auxiliary function $y\mapsto F^{-1}(G(y))$ defined on $%
\mathbb{R}.$ By twice differentiating and using the substitution $q=G(y),$
one finds that it is strictly convex if and only if for $q\in (0,1),$%
\begin{equation}
\frac{g^{\prime }(G^{-1}(q))}{g^{2}(G^{-1}(q))}>\frac{f^{\prime }(F^{-1}(q))%
}{f^{2}(F^{-1}(q))}.  \label{DARACondition2}
\end{equation}%
Since (\ref{DARACondition}) and (\ref{DARACondition2}) are the same, this
last step ends the proof. \hfill $\Box $

\subsection{Proof of Theorem \protect\ref{SUFF2} on page \protect\pageref%
{SUFF2} \label{SUFF2app}}

Assume that the investor has the DARA property. Consider the variable $%
Y_{T}=F^{-1}(G(H_{T})))$ and observe that $Y_{T}\sim F$ (because $G(H_T)\sim%
\mathcal{U}(0,1)$). Since $k(x):=F^{-1}(G(x))$ is strictly convex and
increasing, we only need to prove that $Y_{T}=W_{T}$ (a.s.). $Y_{T}$ is
non-increasing in $\xi _{T}$ and thus cost-efficient. It is thus the a.s.
unique cost-efficient payoff distributed with $F$. Since $W_{T} $ is also
cost-efficient and distributed with $F$ it follows that $Y_{T}=W_{T}$
(a.s.). Conversely, if $W_{T}=k(H_{T})$ then $k(x)=F^{-1}(G(x))$ must hold
where $F$ is the distribution function of $W_{T} $ ($W_{T}$ is
cost-efficient). Since $k(x)$ is strictly convex the DARA property holds
(Theorem \ref{SUFF}). \hfill $\Box $ %

\subsection{Proof of Theorem \protect\ref{SUFF-2} on page \protect\pageref%
{SUFF2} \label{SUFF-2app}}

Since $G(x)$ is the distribution function of a normally distributed random
variable, it is straightforward that ${\mathcal{A}}(x)$ is decreasing if and
only if (\ref{daraBS}) holds true. Furthermore, $F^{-1}(G(x))$ is strictly
convex if and only if $F^{-1}(\Phi (x))$ is strictly convex. The second part
of the theorem now follows from Theorem \ref{SUFF}. \hfill $\Box $ %

\subsection{Proof of Theorem \protect\ref{T2} on page \protect\pageref{T2} 
\label{T2app}}

\proof Denote by $J(x)=1-e^{-\lambda x}$ the distribution of an
exponentially distributed random variable. Since $F^{-1}(\Phi
(x))=F^{-1}(J(J^{-1}(\Phi (x))))$ and $J^{-1}(\Phi (x))$ is strictly convex
(because the exponential distribution corresponds to DARA: see footnote \ref%
{foof}) and strictly increasing, it is sufficient to show that $F^{-1}(J(x))$
is strictly increasing and convex on $\mathbb{R}^{+}\backslash \{0\}$. Since 
$h(x)$ is non-increasing, it follows that $F$ must have support in the form $%
[a,\infty )$ for some $a\in 
\mathbb{R}
$ so that $F$ is strictly increasing on $[a,\infty )$. This implies that $%
F^{-1}(J(x))$ is strictly increasing on $\mathbb{R}^{+}\backslash \{0\}.$
Observe next that $\log (1-F(x))=-\lambda k(x)$ where $k(x):=H^{-1}(F(x))$.
Since $\log(1-F(x))$ is convex and decreasing, $k(x)$ is concave and
increasing i.e., $k^\prime(x)>0$ and $k^{\prime\prime}(x)<0$.
Straight-forward differentiation yields that $k^{-1}(x)=F^{-1}(H(x))$ is
convex, which ends the proof. \hfill $\Box $

\vspace{3cm}

\newpage \singlespacing

\bibliographystyle{plain}
\bibliography{cpv}

\end{document}